\documentclass[a4paper,onecolumn,accepted=2025-11-06]{quantumarticle}
\pdfoutput=1
\usepackage[utf8]{inputenc}
\usepackage[margin=1in]{geometry}

\usepackage{color, xcolor, colortbl}
\usepackage{graphicx,epstopdf}
\usepackage{amsmath,amssymb,amsthm}
\usepackage{algorithm}
\usepackage{algorithmic}
\usepackage{bm}
\usepackage[caption=false]{subfig}
\usepackage{appendix}
\usepackage{multirow}
\usepackage{mathtools}
\usepackage{braket}
\usepackage{hyperref}
\hypersetup{colorlinks=true}
\hypersetup{linkcolor=red}%
\usepackage[english]{babel}
\usepackage[qm]{qcircuit}
\usepackage[capitalize,nameinlink,noabbrev]{cleveref}
\usepackage{tikz}
\usetikzlibrary{shapes.geometric}
\usetikzlibrary{arrows.meta}
\usetikzlibrary{positioning}

\usepackage[sort,square,numbers]{natbib}

\newtheorem{thm}{\protect\theoremname}
  \theoremstyle{plain}
  \newtheorem{lem}[thm]{\protect\lemmaname}
  \theoremstyle{remark}
  
  \theoremstyle{plain}
  \newtheorem*{lem*}{\protect\lemmaname}
  \theoremstyle{plain}
  
  \theoremstyle{plain}
  \newtheorem{cor}[thm]{\protect\corollaryname}
  
  \newtheorem{defn}[thm]{Definition}

\PassOptionsToPackage{USenglish}{babel}
\usepackage[USenglish]{babel}
  \providecommand{\corollaryname}{Corollary}
  \providecommand{\lemmaname}{Lemma}
  \providecommand{\propositionname}{Proposition}
  \providecommand{\remarkname}{Remark}
\providecommand{\theoremname}{Theorem}

\newcommand{\norm}[1]{\|#1\|}

\begin{document}

\title{\centering Quantum algorithms for linear and non-linear fractional reaction-diffusion equations}

\author{Dong An}
\affiliation{Joint Center for Quantum Information and Computer Science, University of Maryland, MD 20742, USA}
\author{Konstantina Trivisa}
\affiliation{Department of Mathematics, University of Maryland, MD 20742, USA}
\affiliation{Institute for Physical Science and Technology, University of Maryland, MD 20742, USA}

\maketitle

\begin{abstract}
High-dimensional fractional reaction-diffusion equations have numerous applications in the fields of biology, chemistry, and physics, and exhibit a range of rich phenomena. While classical algorithms have an exponential complexity in the spatial dimension, a quantum computer can produce a quantum state that encodes the solution with only polynomial complexity, provided that suitable input access is available. In this work, we investigate efficient quantum algorithms for linear and nonlinear fractional reaction-diffusion equations with periodic boundary conditions. For linear equations, we analyze and compare the complexity of various methods, including the second-order Trotter formula, time-marching method, and truncated Dyson series method. We also present a novel algorithm that combines the linear combination of Hamiltonian simulation technique with the interaction picture formalism, resulting in optimal scaling in the spatial dimension. For nonlinear equations, we employ the Carleman linearization method and propose a block-encoding version that is appropriate for the dense matrices that arise from the spatial discretization of fractional reaction-diffusion equations.
\end{abstract}

\tableofcontents

\section{Introduction}
Reaction–diffusion equations arise  in many areas in science and engineering \cite{Britton1986, CantrellCosner2003, Grindrod1996,  Rothe1984, Smoller1994}.  In population dynamics models  in biology, the reaction term typically  accounts for growth, whereas  the diffusion term accounts for migration \cite{Murray2002, NeubertCaswell2000}. The classical diffusion term  has its origin in a model in physics. Recent research investigations indicate that the classical diffusion equation is inadequate to model many real situations, where a particle plume spreads faster than that predicted by the classical model, and may exhibit significant asymmetry \cite{SokolevKlafter20005}. 
In a fractional diffusion equation, the classical Laplace operator in the spatial variable is replaced by a fractional Laplacian of order less than two. 
The fundamental solutions of these equations still exhibit useful scaling properties that make them attractive for applications.

The present article deals with 
\begin{equation}\label{eqn:FRDE}
\begin{split}
    \partial_t u(t,x) &= -(-\Delta)^{\alpha/2} u(t,x) - c(t,x)u(t,x) + a u(t,x) (1-u(t,x)), \quad t \in [0,T], x \in [0,1]^d, \\
    u(0,x) &= u_0(x). 
\end{split}
\end{equation}
Here $(-\Delta)^{\alpha/2}$ is the fractional Laplacian where $0 < \alpha \leq 2$, and $c(t,x)$ is the potential function. 
For the nonlinear term, in this work we only consider the quadratic non-linearity which yields Fisher's equation, but our results can be generalized to equations with high-order polynomial non-linearity. 
For the fractional Laplacian operator, there are several different definitions of $(-\Delta)^{\alpha/2}$ on bounded domain, including spectral definition and Riesz definition~\cite{LischkePangGulianEtAl2020}. 
This works focuses on spectral fractional Laplacian with periodic boundary condition. 
Classical numerical algorithms for solving~\cref{eqn:FRDE} typically require exponential computational resources when the spatial variable is in high dimension: suppose that we use $N$ grid points or basis functions for spatial discretization in each dimension, then the dimension of semi-discretized differential equations becomes as large as $N^d$. 
Therefore we would like to explore the power of quantum algorithms for fractional reaction-diffusion equations and whether quantum algorithms can be efficient in high-dimensional case. 

Quantum algorithms for differential equations aim at preparing a quantum state encoding the solutions at discrete grid points in its amplitudes. 
The first quantum differential equation algorithm was proposed in~\cite{Berry2014}, which transforms differential equations into a linear system of equations using multi-step discretization and then applies quantum linear system algorithms such as HHL algorithm~\cite{HarrowHassidimLloyd2009} or advanced ones~\cite{ChildsKothariSomma2017,SubasiSommaOrsucci2019,AnLin2022,LinTong2020,CostaAnSandersEtAl2022}. 
Since then, there have been remarkable progresses on designing better quantum algorithms, for linear differential equations based on refined discretization~\cite{BerryChildsOstranderEtAl2017,ChildsLiu2020,ChildsLiuOstrander2021,Krovi2022,BerryCosta2022}, time-marching strategy~\cite{FangLinTong2022}, Schr\"odingerization~\cite{JinLiuYu2022}, linear combination of unitaries technique~\cite{AnLiuLin2023}, and for nonlinear differential equations using linearization techniques~\cite{LiuKoldenKroviEtAl2021,AnFangJordanEtAl2022}.
With the caveat that the output is a quantum state encoding solutions in its amplitudes, these quantum algorithms can achieve exponential speedup in the system size compared to classical algorithms. 

However, when directly applied to fractional reaction-diffusion equations, existing generic quantum algorithms are not as efficient as expected due to two major difficulties. 
First, similar as the standard Laplacian operator, fractional Laplacian operator is an unbounded operator, so its spatially discretized version has a huge spectral norm as the spatial dimension and the number of grid points increase. 
Most existing quantum differential equation algorithms scale at least linearly on the spectral norm of the coefficient matrix~\cite{Berry2014,BerryChildsOstranderEtAl2017,ChildsLiu2020,ChildsLiuOstrander2021,Krovi2022,FangLinTong2022,BerryCosta2022,JinLiuYu2022,AnLiuLin2023}, and thus can be computationally expensive in high-dimensional case for accurate simulation. 
Second, when the equation is genuinely fractional (i.e., $\alpha < 2$), the coefficient matrix for the linear part after spatial discretization is unavoidably dense, because fractional differential operators are global operators that depends on the function evaluated in the entire space. 
The dense coefficient matrix poses computational difficulties in solving nonlinear equations. 
This is because all the existing quantum Carleman linearization algorithms require the coefficient matrix to be sparse in order to bypass the difficulty caused by the enlarged Carleman matrix, which is a direct sum of matrices in different dimensions~\cite{LiuKoldenKroviEtAl2021}. 

We remark that the (fractional- or integer-order) Laplacian operator, and more general spatial differential operators, widely appear in various types of partial differential equations. 
In the contexts other than fractional reaction-diffusion equations, there have been several work managing to overcome the computational difficulty brought by its large spectral norm after spatial discretization, but unfortunately those techniques do not apply to the fractional reaction-diffusion equations. 
For example, in real-space Schr\"odinger equation, a poly-logarithmic dependence on the spectral norm of the Laplacian operator can be achieved by simulating the Hamiltonian in the interaction picture~\cite{LowWiebe2019,ChildsLengLiEtAt2022}, which simulates the transformed wavefunction under the rotation associated with the Laplacian operator. 
The resulting interaction picture Hamiltonian becomes bounded and thus can be efficiently simulated. 
To avoid the large spectral norm dependence in the rotations, the algorithm takes advantage of an important feature of the Laplacian operator: it can be diagonalized by the quantum Fourier transform (QFT) circuit and thus can be fastforwarded (i.e., Hamiltonian simulation governed by the Laplacian operator can be implemented with cost independent of evolution time and its spectral norm). 
However, such a technique does not directly work for fractional reaction-diffusion equations. 
The interaction picture transformation requires both forward and backward time evolution of the Laplacian operator. 
This is efficient for Schr\"odinger equations because the dynamics is reversible and both forward and backward time evolution operators are unitary, but fractional reaction-diffusion equation is a dissipative system and implementing its backward time evolution can be prohibitively expensive. 
The recent work~\cite{AnLiuWangZhao2023} proposes efficient quantum algorithms for various partial differential equations beyond Schr\"odinger equations. 
The key technique there is a generalization of the fastforwarding simulation of the Laplacian operator. 
However, the algorithms in~\cite{AnLiuWangZhao2023} require the entire coefficient matrix of the linear part to be fast-forwardable (i.e., its spectral decomposition has quantumly implementable eigenstates and classically computable eigenvalues), which cannot be satisfied even by linear fractional reaction-diffusion equations with the presence of the potential $c(t,x)$. 

\begin{table}[ht]
    \renewcommand{\arraystretch}{1.5}
    \centering
     \scalebox{0.95}{
    \begin{tabular}{c|cccc}\hline
        \multirow{2}{4em}{Method} & \multicolumn{4}{c}{Queries to the matrices}  \\
         \cline{2-5} & $d$ & $\epsilon$ & $T$ & Norm \\\hline 
         Second-order Trotter (\cref{thm:trotter_complexity}) & $\widetilde{\mathcal{O}}(d^{\alpha(1/2+\sigma/2)})$ & $\widetilde{\mathcal{O}}(\epsilon^{-1/2})$ & $\widetilde{\mathcal{O}}(T^{3/2})$ & $\widetilde{\mathcal{O}}((g(T))^{3/2})$ \\
         Time-marching (\cref{thm:time_marching_FRDE}) & $\widetilde{\mathcal{O}}(d^{\alpha(1+2\sigma)})$ & $\mathcal{O}(\text{poly}\log(1/\epsilon))$ & $\widetilde{\mathcal{O}}(T^2)$ & $\widetilde{\mathcal{O}}(Q)$ \\
         Dyson series (\cref{thm:Dyson_method_FRDE}) & $\widetilde{\mathcal{O}}(d^{\alpha(1/2+\sigma)})$ & $\mathcal{O}(\text{poly}\log(1/\epsilon))$ & $\widetilde{\mathcal{O}}(T)$ & $\widetilde{\mathcal{O}}(g(T))$ \\
         LCHS-IP (\cref{thm:LCHS-IP_complexity}) & $\widetilde{\mathcal{O}}(\text{poly}\log(d))$ & $\widetilde{\mathcal{O}}(\epsilon^{-1})$ & $\widetilde{\mathcal{O}}(T)$ & $\widetilde{\mathcal{O}}(g(T)^2)$ \\
         \hline \hline 
         \multirow{2}{4em}{Method} & \multicolumn{4}{c}{Queries to the state preparation}  \\
         \cline{2-5} & $d$ & $\epsilon$ & $T$ & Norm \\\hline 
         Second-order Trotter (\cref{thm:trotter_complexity}) & $\mathcal{O}(1)$ & $\mathcal{O}(1)$ & $\mathcal{O}(1)$ & $\mathcal{O}(g(T))$ \\
         Time-marching (\cref{thm:time_marching_FRDE}) & $\mathcal{O}(1)$ & $\mathcal{O}(1)$ & $\mathcal{O}(1)$ & $\mathcal{O}(Q)$ \\
         Dyson series (\cref{thm:Dyson_method_FRDE}) & $\widetilde{\mathcal{O}}(d^{\alpha(1/2+\sigma)})$ & $\mathcal{O}(\text{poly}\log(1/\epsilon))$ & $\widetilde{\mathcal{O}}(T)$ & $\widetilde{\mathcal{O}}(g(T))$  \\
         LCHS-IP (\cref{thm:LCHS-IP_complexity}) & $\mathcal{O}(1)$ & $\mathcal{O}(1)$ & $\mathcal{O}(1)$ & $\mathcal{O}(g(T))$ \\
         \hline 
    \end{tabular}
    }
    \caption{ Query complexities of differential methods for linear fractional reaction-diffusion equations. 
    Here $d$ is the spatial dimension, $\epsilon$ is the tolerated error in $2$-norm, $T$ is the evolution time, $\alpha$ is the half order of the fractional Laplacian operator ranging in $(0,2]$ and $\sigma$ is the parameter of the Gevrey class defined in~\cref{eqn:gevrey_class}. 
    The function $g(T) \geq \|\vec{u}(0)\|/\|\vec{u}(T)\|$ describes the decay of the spatially discretized solution, $Q$ is the decay corrected by the spectral norm of infinitesimal evolution operators as defined in~\cref{eqn:time_marching_Q_def}, and we always have $Q \leq g(T)$. }
    \label{tab:comp_linear_FRDE}
\end{table}

In this work, we investigate efficient quantum algorithms for fractional reaction-diffusion equations. 
The majority of our work is devoted to linear fractional reaction-diffusion equations. 
We numerically treat the equations by the method of lines, i.e., first discretizing the spatial variable to obtain a system of ODEs
\begin{equation}\label{eqn:FRDE_ODE_intro}
    \frac{d}{dt}\vec{u} = -B \vec{u} - C(t) \vec{u}, 
\end{equation}
and then solving the resulting ODE system with different quantum ODE algorithms. 
Here $\vec{u}$ represents the solution evaluated at different spatial grid points, $B$ is the discretized fractional Laplacian operator, and $C(t)$ is the potential matrix. 
For time evolution, we consider four different algorithms: second-order Trotter formula~\cite{ChildsSuTranEtAl2020}, time-marching method~\cite{FangLinTong2022}, truncated Dyson series method~\cite{BerryCosta2022}, and linear combination of Hamiltonian simulation in the interaction picture (LCHS-IP), which is a novel method that combines the techniques in the two references~\cite{AnLiuLin2023,LowWiebe2019}. 
We analyze the complexity of these four methods and the results are shown in~\cref{tab:comp_linear_FRDE}. 
Our main results and contributions are summarized as follows: 
\begin{enumerate}
    \item Second-order Trotter formula: unlike Hamiltonian simulation,~\cref{eqn:FRDE_ODE_intro} is not a unitary dynamics, and in the Trotter formula we need to implement non-unitary operators $e^{-Bs}$ and $e^{-C(t)s}$. 
    We discuss efficient construction of these operators via controlled rotations, and implementing their multiplication through a generalization of the compression gadget technique~\cite{LowWiebe2019,FangLinTong2022} with only poly-logarithmic many ancilla qubits. 
    For complexity analysis, we derive an improved Trotter error bound that avoids the exponential factor in~\cite{ChildsSuTranEtAl2020} and generalizes to the time-dependent case. 
    Compared to time-marching and truncated Dyson series methods, second-order Trotter has better dependence on the dimension $d$ thanks to its commutator scalings, but has worse dependence on the precision. 
    \item Time-marching method: we directly apply the standard time-marching method in~\cite{FangLinTong2022} and analyze its complexity for fractional reaction-diffusion equations. 
    It has low state preparation cost and poly-logarithmic dependence on precision, but has worse scalings in the dimension $d$ and the evolution time $T$. 
    \item Truncated Dyson series method: we directly apply the standard truncated Dyson series method in~\cite{BerryCosta2022} and analyze its complexity for fractional reaction-diffusion equations. 
    It still depends polynomially on the dimension due to its spectral norm dependence and has high state preparation cost due to the usage of quantum linear system algorithms, but can achieve poly-logarithmic scaling in the precision and linear scaling in time simultaneously. 
    \item LCHS-IP: The LCHS-IP method is a novel method that combines the linear combination of Hamiltonian simulation (LCHS)~\cite{AnLiuLin2023} technique with the interaction picture Hamiltonian simulation~\cite{LowWiebe2019}. 
    The LCHS method first represents the evolution operator of~\cref{eqn:FRDE_ODE_intro} as a linear combination of several Hamiltonian simulation problems associated with the matrices $B$ and $C(t)$. 
    To avoid the computational overhead brought by the discretized fractional Laplacian $B$, we implement each Hamiltonian simulation in the interaction picture by rotating the Hamiltonian with respect to $B$. 
    Therefore, the resulting algorithm only has poly-logarithm dependence on the dimension $d$ and thus is the most preferable algorithm in the high-dimensional case. 
    It also has low state preparation cost, but only linear scaling in precision. 
\end{enumerate}

For nonlinear fractional reaction-diffusion equations, we discuss a block-encoding version of the Carleman linearization technique to deal with the dense coefficient matrix. 
Let $\vec{u}(t)$ denote the solution vector of the fractional reaction-diffusion equation after spatial discretization. 
Standard quantum Carleman linearization algorithm~\cite{LiuKoldenKroviEtAl2021} considers the dynamics of the enlarged vector $[\vec{u}(t);\vec{u}(t)^{\otimes 2}; \cdots; \vec{u}(t)^{\otimes M}]$, which approximately satisfies a linear system of ODEs governed by a so-called Carleman matrix. 
Since $\vec{u}(t)^{\otimes m}$'s are in different sizes for different powers, the Carleman matrix is a direct sum of matrices in different dimensions. 
In our work, we propose a simple generalization of the Carleman linearization, by extending the Carleman matrix to even higher dimension such that it becomes the direct sum of matrices in the same dimension and the corresponding solution vector $[\vec{w}_1;\vec{w}_2; \cdots; \vec{w}_M]$ can be exactly mapped to the solution $[\vec{u}(t);\vec{u}(t)^{\otimes 2}; \cdots; \vec{u}(t)^{\otimes M}]$ via discarding all the zero entries. 
Then the extended Carleman matrix can be easily block-encoded through the block-encoding of the original coefficient matrix, and thus the corresponding extended linearized system can be solved via quantum linear differential algorithms with block-encoding input model (e.g.,~\cite{FangLinTong2022,BerryCosta2022,AnLiuLin2023}). 
This makes the quantum Carleman linearization algorithm applicable to the equations with dense coefficient matrices, including the fractional reaction-diffusion equations.

The rest of this paper is organized as follows. 
\cref{sec:preliminaries} first discusses the mathematical setup of the fractional reaction-diffusion equations, notations being used throughout the paper and some preliminary results. 
The main results of this paper start by~\cref{sec:linear_no_potential} with the simplest case where the equations only involve a single fractional Laplacian operator without potential or nonlinear term, then~\cref{sec:linear} considers the general linear equations with potential. 
Quantum algorithm for nonlinear equations is presented in~\cref{sec:nonlinear}, followed by conclusion and open questions in~\cref{sec:conclusion}.

\section{Preliminaries}\label{sec:preliminaries}

We start with a more rigorous setup of the fractional reaction-diffusion equation we are interested in and a summary of theoretical tools and technical lemmas being used in our analysis. 

\subsection{Setup}

We consider the spatial fractional reaction-diffusion equations~\cref{eqn:FRDE}. 
Let $N$ be a positive integer and we discretize the spatial variable $x$ using equi-distant nodes $(j/N)$, $j \in [N]^d$. 
Quantum algorithms for solving~\cref{eqn:FRDE} aim at preparing a quantum state approximately encoding $u(T,j/N)$ in its amplitude, \emph{i.e.}, $\frac{1}{\|(u(T,j/N))_{j\in[N]^d}\| } \sum_{j\in [N]^d} u(T,j/N) \ket{j}$. 

There are several different definitions of $(-\Delta)^{\alpha/2}$ on bounded domain (see the recent paper~\cite{LischkePangGulianEtAl2020} for a comprehensive review). 
The \emph{spectral fractional Laplacian} is defined using the eigenvalues and eigenfunctions of the original Laplacian $(-\Delta)$. 
Suppose that $\lambda_j$'s are the eigenvalues of $(-\Delta)$ and $e_j(x)$'s are the corresponding eigenfunctions, then the spectral fractional Laplacian is defined to be 
\begin{equation}\label{eqn:def_spectral_Laplacian}
    (-\Delta)^{\alpha/2} v(x) = \sum_{j} \lambda_j^{\alpha/2} (v,e_j)_{L^2} e_j(x) 
\end{equation}
where $(\cdot,\cdot)_{L^2}$ denotes the $L^2$ inner product on $[0,1]^d$. 
An alternative definition is the \emph{Riesz fractional Laplacian} 
\begin{equation}\label{eqn:def_Riesz_Laplacian}
    (-\Delta)^{\alpha/2} v(x) = \frac{2^{\alpha} \Gamma(\frac{\alpha}{2} + \frac{d}{2})}{\pi^{d/2} |\Gamma(-\frac{\alpha}{2})| }  \text{p.v.} \int_{\mathbb{R}^d} \frac{v(x) - v(y)}{|x-y|^{d+\alpha}} dy. 
\end{equation}
Here $\Gamma(s)$ denotes the gamma function and p.v. refers to the principle value integral.  
Notice that the definition~\cref{eqn:def_Riesz_Laplacian} is not closed yet since the integral is over the entire space $\mathbb{R}^d$ while $v(x)$ is confined in the cube $[0,1]^d$. 
Therefore we must enforce boundary condition on $\mathbb{R}^d\setminus[0,1]^d$, \emph{e.g.}, the homogeneous Dirichlet boundary condition $v(x) = 0$ for all $x \in \mathbb{R}^d\setminus[0,1]^d$. 
For periodic boundary condition, the definition is unambiguous that we will always follow~\cref{eqn:def_spectral_Laplacian}, while either definition is commonly used with other boundary conditions. 
In this work, we will focus on the periodic boundary condition and use the spectral definition for the fractional Laplacian operator.

\subsection{Notations}

We use $\vec{a}$ (a letter with an array above) to denote a possibly unnormalized vector. 
When $u(t,x)$ refers to a scalar-valued function with arguments $t \in \mathbb{R}$ and $x \in [0,1]^d$, we use $(u(t,j/N))_{j\in [N]^d}$ or simply $(u(t,j/N))$ to denote the $N^d$-dimensional vector with entries $u(t,j/N)$, where $j \in [N]^d$ and $[N] = \left\{0,1,\cdots,N-1\right\}$ for a positive integer $N$. 
In our analysis, we also use $\vec{u}_0$ as a shorthand notation of the initial condition vector $(u(0,j/N))$, and $\vec{u}(t)$ as $(u(t,j/N))$ for a fixed time $t$. 

For a vector $\vec{a}$, we use $\|\vec{a}\|$ without subscript for the standard vector 2-norm of $\vec{a}$, and $\|\vec{a}\|_p$ for vector $p$-norm.  
The notation $\ket{\vec{a}}$ with ket notation denotesthe corresponding quantum state, \emph{i.e.}, the normalized vector $\vec{a}/\|\vec{a}\|$. 

Let $A$ and $B$ be two matrices. 
We use $[A,B]$ to denote the commutator between $A$ and $B$, defined as $AB-BA$. 
For a matrix $A$, $\|A\|$ denotes its spectral norm or equivalently the matrix 2-norm.

\subsection{Quantum linear algebra}

To design and analyze quantum algorithms for the spatial fractional reaction-diffusion equations, we need to frequently implement linear algebra operations, including matrix-vector multiplication, matrix-matrix addition and multiplication. 
To this end, we briefly introduce the concept and properties of block-encoding, which is a widely-used quantum input model for possibly non-unitary matrix. 

We start with the definition of block-encoding. 

\begin{defn}[Block-encoding]
    Suppose that $A$ is a $2^s$-dimensional matrix, then we say that the $(s+n)$-qubit unitary $U$ is an $(\alpha,n,\epsilon)$-block-encoding of $A$, if 
    \begin{equation}
        \left\| A - \alpha \left( (\bra{0}^{\otimes n} \otimes I) U (\ket{0}^{\otimes n} \otimes I)\right)  \right\| \leq \epsilon. 
    \end{equation}
\end{defn}

Intuitively, in the block-encoding, a matrix is represented as the upper-left block of a unitary matrix as 
\begin{equation}
    U \approx \left(\begin{array}{cc}
        A/\alpha & * \\
        * & *
    \end{array}\right). 
\end{equation}
$\alpha$ is called the block-encoding factor such that $\alpha \geq \|A\|$, since the larger matrix $U$ is supposed to be unitary and the norm of its sub-block should be bounded by $1$. 

With block-encoding structure, we may implement matrix linear algebra operations. 
For example, matrix addition can be directly implemented by the linear combination of unitaries (LCU) technique, which has become an important subroutine in designing various quantum algorithms including Hamiltonian simulation~\cite{ChildsWiebe2012}, solving linear systems~\cite{ChildsKothariSomma2017} and differential equations~\cite{AnLiuLin2023}. 
Here we present the LCU lemma from~\cite{GilyenSuLowEtAl2019}. 

\begin{lem}\label{lem:LCU}
    Let $A = \sum_{j=0}^{m-1} y_j A_j$ and $\|y\|_1 \leq \beta$. 
    Suppose that $(P_L,P_R)$ is a pair of $b$-qubit unitaries such that $P_L \ket{0} \sum_{j=0}^{m-1} c_j\ket{j}$ and $P_R \ket{0} \sum_{j=0}^{m-1} d_j\ket{j}$ with $\sum_{j=0}^{m-1} |\beta (c_j^* d_j - y_j)| < \epsilon_1$, and $W = \sum_{j=0}^{m-1} \ket{j}\bra{j} \otimes U_j$ where $U_j$ is a $(\alpha,a,\epsilon_2)$-block-encoding of $A_j$. 
    Then we can construct a $(\alpha\beta,a+b,\alpha\epsilon_1+\alpha\beta\epsilon_2)$-block-encoding of $A$ with a single use of $W$, $P_L^{\dagger}$ and $P_R$. 
\end{lem}

Matrix multiplication can also be implemented via block-encodings. 
For two arbitrary matrices $A$ and $B$, a straightforward approach of constructing the block-encoding of $AB$ is to multiply together the block-encodings of $A$ and $B$, as shown in the following lemma~\cite{GilyenSuLowEtAl2019}. 

\begin{lem}
    If $U_A$ is an $(\alpha_A,n_A,\epsilon_A)$-block-encoding of $A$, and $U_B$ is an $(\alpha_B,n_B,\epsilon_B)$-block-encoding of $B$, then $(I_{n_B}\otimes U_A)(I_{n_A}\otimes U_B)$ is an $(\alpha_A\alpha_B,n_A+n_B,\alpha_A\epsilon_B + \alpha_B\epsilon_A)$-block-encoding of $AB$. 
\end{lem}

\begin{figure}
    \centerline{
    \Qcircuit @R=1em @C=1em {
    \text{Counter} \quad\quad\quad\quad\quad & \gate{\text{ADD}^J} & \gate{\text{ADD}^{\dagger}} & \qw & \gate{\text{ADD}^{\dagger}} & \qw & \cdots & \quad & \qw & \gate{\text{ADD}^{\dagger}} & \qw \\
    \text{Ancilla} \quad\quad\quad\quad\quad & \multigate{1}{U_0} & \ctrlo{-1} & \multigate{1}{U_1} & \ctrlo{-1} & \qw & \cdots & \quad & \multigate{1}{U_{J-1}} & \ctrlo{-1} & \qw \\ 
    \text{System}  \quad\quad\quad\quad\quad & \ghost{U_0} & \qw & \ghost{U_1} & \qw & \qw & \cdots & \quad & \ghost{U_{J-1}} & \qw & \qw
    }
    }
    \caption{ Quantum circuit for compression gadget to block encode $A_{J-1}\cdots A_1A_0$. Here the Counter register contains $\lceil\log_2 J\rceil + 1$ qubits and the ancilla register contains $\max n_j$ qubits. ADD implements addition by $1$ modulo $2^{\lceil\log_2 J\rceil + 1}$, and $U_j$ is an $(\alpha_j,n_j,0)$-block-encoding of the matrix $A_j$. }
    \label{fig:compression_gadget}
\end{figure}
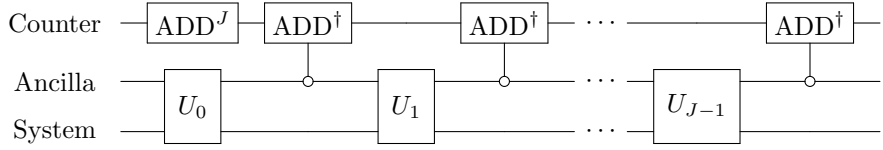

Despite its simplicity, such a straightforward approach for matrix multiplication may incur large space overhead. 
This is because we need to enlarge the ancilla register at each step of the multiplication, so the multiplication of $J$ many matrices will in general require $\mathcal{O}(J)$ ancilla qubits. 
To overcome this issue,~\cite{LowWiebe2019} introduces a technique called compression gadget, which is further simplified in~\cite{FangLinTong2022}. 
The circuit is given in~\cref{fig:compression_gadget}, and the idea of compression gadget is to use a counter register to keep track of the multiplication in a coherent way. 
Specifically, let ADD implements addition by $1$ modulo the 2-power of the number of the qubits in the counter register. 
One can first apply ADD to lift the counter register from $0$ to $L-1$. 
Then we sequentially apply the block-encodings using the same ancilla register, but after each block-encoding we apply a controlled $\text{ADD}^{\dagger}$ to reduce the counter by $1$ if the corresponding block-encoding was successfully implemented. 
Therefore, a $0$ final outcome of the counter register implies successful application of all the block-encodings. 
We present the result in~\cite{FangLinTong2022}, in which interested readers may find more details on its proof. 

\begin{lem}\label{lem:block_encoding_product}
    For $0 \leq j \leq J-1$, let $U_{j}$ be an $(\alpha_j,n_j,0)$-block-encoding of $A_j$. 
    Then an $(\alpha,n,0)$-block-encoding of $A_{J-1} \cdots A_1 A_0$ can be constructed using one application of each $U_j$, where $\alpha = \alpha_0\alpha_1\cdots \alpha_{J-1}$ and $n = \max n_j + \lceil \log_2(J) \rceil + 1$. 
\end{lem}

\section{Linear equations without potential}\label{sec:linear_no_potential}

We start with the simplest case where $c(t,x) = 0$ and $a = 0$. 
In this case, the right hand side of~\cref{eqn:FRDE} only involves a fractional Laplacian operator. 
By the spectral definition, this operator has closed-form eigenvalues and known eigenfunctions, so the corresponding time-evolution operator can be implemented fast-forwardly~\cite{AnLiuWangZhao2023} (\emph{i.e.}, query complexity is independent of the spectral norm and the evolution time). 
Here we present our algorithm in general high dimension and establish rigorous complexity estimate taking into consideration the spatial discretization errors.

We consider general $d$-dimensional case. 
The eigenvalues and eigenfunctions of $(-\Delta)$ are $4\pi^2(k_0^2+\cdots+k_{d-1}^2)$ and $e^{2\pi i (k_0x_0+\cdots + k_{d-1}x_{d-1})}$, where $x = (x_0,\cdots,x_{d-1})$ denotes the spatial coordinate and $k = (k_0,\cdots,k_{d-1})$ denotes a set of integers. 
Let the Fourier series of $u_0(x)$ be 
\begin{equation}
    u_0(x) = \sum_{k\in \mathbb{Z}^d} \hat{u}_k e^{2\pi i (k_0x_0+\cdots + k_{d-1}x_{d-1})}, 
\end{equation}
where $\hat{u}_k$ denotes the Fourier coefficients. 
Then the solution of~\cref{eqn:FRDE} has the form 
\begin{equation}\label{eqn:solu_linear_high_d}
    u(T,x) = \sum_{k\in \mathbb{Z}^d} \hat{u}_k e^{-(4\pi^2(k_0^2+\cdots+k_{d-1}^2))^{\alpha/2} T} e^{2\pi i (k_0x_0+\cdots + k_{d-1}x_{d-1})}. 
\end{equation}
Numerical solutions can be obtained by truncating the Fourier series at a finite order.

\cref{eqn:solu_linear_high_d} can be quantumly implemented using quantum Fourier transform (QFT) and controlled rotations. 
Recall that our goal is to prepare an approximation of the quantum state encoding the normalized solution at discrete spatial grid points $(j/N)_{j \in [N]^d}$. 
Let $N$ be the number of the grid points in each spatial dimension. 
Suppose that we are given the oracle $O_{u_0}$ that prepares the normalized initial condition 
\begin{equation}
    \ket{u_0} = \frac{1}{\|\vec{u}_0\|}\sum_{n\in [N]^d} u_0(n_0/N,\cdots,n_{d-1}/N)\ket{n_0}\cdots\ket{n_{d-1}}, 
\end{equation}
where 
\begin{equation}
    \vec{u}_0 = \sum_{n\in [N]^d} u_0(n_0/N,\cdots,n_{d-1}/N)\ket{n_0}\cdots\ket{n_{d-1}}. 
\end{equation}
We first compute a quantum state encoding the Fourier coefficients $\hat{u}_k$ by QFT. 
Specifically, let $\omega_N = e^{2\pi i/N}$ and $\mathcal{F}$ denote the one-dimensional QFT, \emph{i.e.}, for any computational basis state $\ket{j}$, 
\begin{equation}\label{eqn:def_QFT}
    \mathcal{F} \ket{j} = \frac{1}{\sqrt{N}} \sum_{l=0}^{N-1} \omega_N^{jl} \ket{l}. 
\end{equation}
Then 
\begin{align}
    (\mathcal{F}^{-1})^{\otimes d} \vec{u}_0 &= \frac{1}{N^{d/2}} \sum_{n\in [N]^d} u_0(n_0/N,\cdots,n_{d-1}/N)\left(\sum_{m_0=0}^{N-1} \omega_N^{-n_0m_0} \ket{m_0}\right) \cdots\left(\sum_{m_{d-1}=0}^{N-1} \omega_N^{-n_{d-1}m_{d-1}} \ket{m_{d-1}}\right) \\
    &= \frac{1}{N^{d/2}} \sum_{m\in [N]^d} \sum_{n\in [N]^d} u_0(n_0/N,\cdots,n_{d-1}/N)\omega_N^{-n_{0}m_0-\cdots-n_{d-1}m_{d-1}} \ket{m_0}\cdots\ket{m_{d-1}} \\
    &= \frac{1}{N^{d/2}} \sum_{m\in [N]^d} \sum_{n\in [N]^d} \sum_{k\in \mathbb{Z}^d} \hat{u}_k e^{2\pi i (k_0n_0/N+\cdots + k_{d-1}n_{d-1}/N)} \omega_N^{-n_0m_0-\cdots-n_{d-1}m_{d-1}} \ket{m_0}\cdots\ket{m_{d-1}} \\
    & = \frac{1}{N^{d/2}} \sum_{m\in [N]^d} \sum_{k\in \mathbb{Z}^d} \hat{u}_k \left(\sum_{n_0=0}^{N-1} \omega_N^{n_0(k_0-m_0)}\right)\cdots \left(\sum_{n_{d-1}=0}^{N-1} \omega_N^{n_{d-1}(k_{d-1}-m_{d-1})}\right)  \ket{m_0}\cdots\ket{m_{d-1}}\\
    &= N^{d/2} \sum_{m\in [N]^d} \sum_{j\in \mathbb{Z}^d} \hat{u}_{m+jN}  \ket{m_0}\cdots\ket{m_{d-1}}. 
\end{align}
When the function $u_0(x)$ satisfies certain regularity assumption 
(i.e., at least $(d+2)$-th order continuously differentiable specified in~\cref{thm:fractional_heat}), 
the summation $\sum_{j\in \mathbb{Z}^d} \hat{u}_{m+jN} $ is dominated by the index with smallest absolute value, because the Fourier coefficients decay rapidly with respect to its frequency. 
Therefore 
\begin{equation}\label{eqn:DFT_calculation_high_d}
    (\mathcal{F}^{-1})^{\otimes d} \ket{\vec{u}_0} \approx \frac{N^{d/2}}{\|\vec{u}_0\|} \sum_{m\in [N]^d} \hat{u}_{i(m)}  \ket{m_0}\cdots\ket{m_{d-1}}, 
\end{equation}
where $i(m) = (i_0(m),\cdots,i_{d-1}(m))$ represents an $d$-dimensional vector with the $j$-th index
\begin{equation}\label{eqn:def_i_j}
    i_j(m) = \begin{cases} 
    m_j, & \text{ if } 0 \leq m_j \leq N/2\\
    m_j-N, & \text{ if } N/2+1 \leq m_j \leq N-1. 
    \end{cases}
\end{equation}

Now we append three ancilla registers to~\cref{eqn:DFT_calculation_high_d} and get the (approximate) quantum state 
\begin{equation}
    \frac{N^{d/2}}{\|\vec{u}_0\|} \sum_{m\in [N]^d} \hat{u}_{i(m)}  \ket{m_0}\cdots\ket{m_{d-1}}\ket{0}\ket{0}\ket{0}. 
\end{equation}
The first two ancilla registers are used to binarily encode $e^{-(4\pi^2(i_0(m)^2+\cdots+i_{d-1}(m)^2))^{\alpha/2} T}$ and the third register is used for rotation. 
Specifically, suppose that we are given the oracle that encodes the eigenvalues as 
\begin{equation}\label{eqn:def_O_high_d_linear}
    O_1: \ket{m_0}\cdots\ket{m_{d-1}}\ket{0} \rightarrow \ket{m_0}\cdots\ket{m_{d-1}}\ket{(2\pi\|i(m)\|)^{\alpha}}, 
\end{equation}
and the oracle for computing an exponential function as 
\begin{equation}\label{eqn:def_O_exp_1}
    O_{\exp,1} : \ket{x}\ket{0} \rightarrow \ket{x}\ket{e^{-x T}}. 
\end{equation}
Notice that both functions have closed-form expression so the oracles can be efficiently constructed using classical arithmetic~\cite{NielsenChuang2000}. 
Applying $O_1$ on the index and first ancilla registers and then $O_{\exp,1}$ on the first and second ancilla registers yields 
\begin{equation}\label{eqn:DFT_calculation_high_d_2}
    \frac{N^{d/2}}{\|\vec{u}_0\|} \sum_{m\in [N]^d} \hat{u}_{i(m)}  \ket{m_0}\cdots\ket{m_{d-1}}\ket{\|i(m)\|}\ket{e^{-(2\pi \|i(m)\|)^{\alpha}T}}\ket{0}. 
\end{equation}
Let $\text{c-}R$ denote the controlled rotation
\begin{equation}\label{eqn:def_cR_rotation}
    \text{c-}R: \ket{\theta} \ket{0} \rightarrow \ket{\theta} \left(\theta \ket{0} + \sqrt{1-|\theta|^2}\ket{1} \right). 
\end{equation}
Then, by applying $\text{c-}R$ on the last two registers in~\cref{eqn:DFT_calculation_high_d_2}, we obtain 
\begin{equation}
    \frac{N^{d/2}}{\|\vec{u}_0\|} \sum_{m\in [N]^d} \hat{u}_{i(m)} e^{-(2\pi \|i(m)\|)^{\alpha}T} \ket{m_0}\cdots\ket{m_{d-1}}\ket{\|i(m)\|}\ket{e^{-(2\pi \|i(m)\|)^{\alpha}T}}\ket{0} + \ket{\perp}, 
\end{equation}
where $\ket{\perp}$ represents the orthogonal part with $1$ in the last ancilla register. 
Uncomputing the first two ancilla registers yields 
\begin{equation}\label{eqn:state_1_high_d}
    \frac{N^{d/2}}{\|\vec{u}_0\|} \sum_{m\in [N]^d} \hat{u}_{i(m)} e^{-(2\pi \|i(m)\|)^{\alpha}T} \ket{m_0}\cdots\ket{m_{d-1}}\ket{0}\ket{0}\ket{0} + \ket{\perp}, 
\end{equation}

Notice that, by replacing $u_0(x)$ with $u(T,x)$ in~\cref{eqn:DFT_calculation_high_d} and using~\cref{eqn:solu_linear_high_d}, we have 
\begin{equation}
    (\mathcal{F}^{-1})^{\otimes d} \vec{u}(T,x) \approx N^{d/2} \sum_{m\in [N]^d} \hat{u}_{i(m)} e^{-(2\pi\|i(m)\|)^{\alpha} T} \ket{m_0}\cdots\ket{m_{d-1}}. 
\end{equation}
Therefore, applying the QFT $\mathcal{F}^{\otimes d}$ to~\cref{eqn:state_1_high_d} approximately gives 
\begin{equation}
    \frac{1}{\|\vec{u}_0\|} \sum_{n \in [N]^d } u(T,n/N) \ket{n_0}\cdots\ket{n_{d-1}}\ket{0}\ket{0}\ket{0} + \ket{\perp}. 
\end{equation}
Measuring the ancilla registers to get all $0$'s yields an approximation of $\ket{u(T)}$, and the averaged number of repeats for success after amplitude amplification scales $\mathcal{O}(\|\vec{u}_0\|/\|\vec{u}(T)\|)$. 

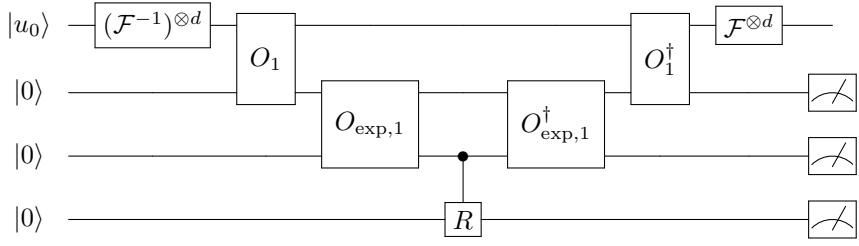
\begin{figure}
    \centerline{
    \Qcircuit @R=1em @C=1em {
    \ket{u_0} \quad\quad\quad & \gate{(\mathcal{F}^{-1})^{\otimes d}} & \multigate{1}{O_1} & \qw & \qw & \qw & \multigate{1}{O_1^{\dagger}} & \gate{\mathcal{F}^{\otimes d}} & \qw  \\
    \ket{0} \quad\quad\quad & \qw & \ghost{O_1} & \multigate{1}{O_{\exp,1}} & \qw & \multigate{1}{O_{\exp,1}^{\dagger}} & \ghost{O_1^{\dagger}} & \qw & \meter\\ 
     \ket{0} \quad\quad\quad & \qw & \qw & \ghost{O_{\exp,1}} & \ctrl{1} & \ghost{O_{\exp,1}^{\dagger}} & \qw & \qw & \meter\\ 
    \ket{0} \quad\quad\quad & \qw & \qw & \qw & \gate{R} & \qw & \qw & \qw & \meter \\
    }
    }
    \caption{ Quantum circuit for solving linear fractional reaction-diffusion equations without potential. Here $O_1$, $O_{\exp,1}$ are the oracles defined in~\cref{eqn:def_O_high_d_linear} and~\cref{eqn:def_O_exp_1}, and $R$ is the single-qubit rotation gate defined in~\cref{eqn:def_cR_rotation}. }
    \label{fig:circuit_FRDE_linear}
\end{figure}

We summarize the quantum algorithm in~\cref{fig:circuit_FRDE_linear} and present its error and complexity estimates in the following result. 
The main result is that the overall error will decrease as the number of the grid points increases, since more grid points imply larger truncation order in the Fourier series. 
Detailed proof can be found in~\cref{app:proof_linear_without_potential}. 

\begin{thm}\label{thm:fractional_heat}
    Consider solving~\cref{eqn:FRDE} with $a = c = 0$. 
    Suppose that we are given the state preparation oracle $O_{u_0}$ and the oracles $O_1$, $O_{\exp,1}$ as defined in~\cref{eqn:def_O_high_d_linear} and~\cref{eqn:def_O_exp_1}. 
    Furthermore, suppose that there exists an integer $p \geq d + 2$ such that $u(t,x)$ is $p$-th order spatially continuously differentiable. 
    Then there exists a quantum algorithm that prepares an approximation of the quantum state $\ket{\vec{u}(T)}$ with success probability $\Omega(1)$\footnote{The $\Omega(1)$ here is referred to all variables, i.e., an absolute constant that is smaller than $1$ (e.g., $0.9$ or $0.99$), so the success probability is bounded from below by it (e.g., $0.9$ or $0.99$).} and $2$-norm error at most 
    \begin{equation}
        \frac{4 \max_{j,t\in\left\{0,T\right\}}\|\partial_{x_j}^p u(t,x) \|_{L^1}}{\|\vec{u}(T)\|(\pi/2)^p N^{p-d} }, 
    \end{equation}
    using $\mathcal{O}(\|\vec{u}_0\|/\|\vec{u}(T)\|)$ queries to $O_{u_0}, O_1, O_{\exp,1}$, their inverse and controlled versions, and \\ $\mathcal{O}((\|\vec{u}_0\|/\|\vec{u}(T)\|) d\log^2(N))$ additional gates. 
\end{thm}

\section{Linear equations with potential}\label{sec:linear}

Consider the linear spatial fractional reaction-diffusion equations 
\begin{equation}\label{eqn:FRDE_linear}
\begin{split}
    \partial_t u(t,x) &= -(-\Delta)^{\alpha/2} u(t,x) - c(t,x) u(t,x) , \quad t \in [0,T], x \in [0,1]^d, \\
    u(0,x) &= u_0(x). 
\end{split}
\end{equation}
Throughout this section, we assume the potential function $c(t,x)$ to be non-negative. 
We will first discuss the validity of this assumption by so-called shifting equivalence. 
The equation~\cref{eqn:FRDE_linear} can then be treated by the method of lines, \emph{i.e.}, first discretizing the spatial variable to obtain a system of ODEs, and then solving the ODE with different quantum algorithms. 
We consider four different quantum algorithms: second-order Trotter formula, time-marching method, truncated Dyson series method, and linear combination of Hamiltonian simulation in the interaction picture.

\subsection{Shifting equivalence}

We consider the shifted PDE 
\begin{equation}\label{eqn:FRDE_linear_shifted}
\begin{split}
    \partial_t v(t,x) &= -(-\Delta)^{\alpha/2} v(t,x) - \widetilde{c}(t,x) v(t,x) , \quad t \in [0,T], x \in [0,1]^d, \\
    v(0,x) &= u_0(x). 
\end{split}
\end{equation}
Here 
\begin{equation}
    \widetilde{c}(t,x) = c(t,x) - \gamma(t)
\end{equation}
where $\gamma(t)$ is a real scalar-valued function. 
An important observation is that~\cref{eqn:FRDE_linear} and~\cref{eqn:FRDE_linear_shifted} are \emph{quantumly} equivalent in the sense that the normalized solutions are the same. 
Specifically, let $u$ denote the solution of~\cref{eqn:FRDE_linear}, then 
\begin{equation}
\begin{split}
    \partial_t(e^{\int_0^t \gamma(s)ds } u) &= e^{\int_0^t \gamma(s)ds}  \partial_t u + \gamma(t) e^{\int_0^t \gamma(s)ds } u  \\
    & =e^{\int_0^t \gamma(s)ds}   \left( -(-\Delta)^{\alpha/2} u- c(t,x) u  \right) + \gamma(t) e^{\int_0^t \gamma(s)ds } u  \\
    & = -(-\Delta)^{\alpha/2} (e^{\int_0^t \gamma(s)ds } u) - \widetilde{c}(t,x) (e^{\int_0^t \gamma(s)ds} u). 
\end{split}
\end{equation}
Therefore $e^{\int_0^t \gamma(s)ds} u$ solves~\cref{eqn:FRDE_linear_shifted} and only differs $u$ by a multiplicative constant factor at fixed time, which implies that they are the same in a quantum state representation, \emph{i.e.}, 
\begin{equation}
    \frac{1}{\|\vec{u}\|} \sum_{n\in[N]^d} u(t,n/N)\ket{n} = \frac{1}{\|e^{\int_0^t \gamma(s)ds}\vec{u}\|} \sum_{n\in[N]^d} e^{\int_0^t \gamma(s)ds}u(t,n/N)\ket{n}. 
\end{equation}

We will choose the shifting parameter to be 
\begin{equation}
    \gamma(t) = \min_{x} c(t,x). 
\end{equation}
Therefore the function $\widetilde{c}(t,x)$ is non-negative. 
For notation simplicity, we will directly assume that the original $c(t,x)$ in~\cref{eqn:FRDE_linear} to be non-negative instead of introducing new notations with tildes.

\subsection{Spatial discretization}

Our goal is to prepare a quantum state encoding $u(T,x)$ at equi-distant grid points $(n_0/N,\cdots,n_{d-1}/N)$ where $n_j \in [N]$. 
Motivated by the spectral decomposition, we define 
\begin{equation}\label{eqn:def_B}
    B =  (\mathcal{F})^{\otimes d}  D (\mathcal{F}^{-1})^{\otimes d} 
\end{equation}
where $\mathcal{F}$ is the one-dimensional quantum Fourier transform defined in~\cref{eqn:def_QFT}, and $D$ is an $N^d$-dimensional diagonal matrix 
\begin{equation}\label{eqn:def_D}
    D = \sum_{n \in [N]^d} 2^{\alpha}\pi^{\alpha} \|i(n)\|^{\alpha}  \ket{n_0}\cdots \ket{n_{d-1}} \bra{n_0}\cdots \bra{n_{d-1}}. 
\end{equation}
For each $t$, let $C(t)$ be an $N^d$-dimensional diagonal matrix 
\begin{equation}
    C(t) = \sum_{n \in [N]^d} c(t,n/N) \ket{n_0}\cdots \ket{n_{d-1}} \bra{n_0}\cdots \bra{n_{d-1}}. 
\end{equation}
Then we consider the spatially discretized equation 
\begin{equation}\label{eqn:FRDE_linear_ODE_shifted}
\begin{split}
    \frac{d}{dt} \vec{u} &= -B \vec{u} - C(t) \vec{u} , \quad t \in [0,T] \\
    \vec{u}(0) &= \vec{u}_0. 
\end{split}
\end{equation}

In order to bound the spatial discretization error by $\epsilon$, we need to choose a sufficiently large $N$. 
In the following, we derive an error bound in terms of $N$. 
The proof of this result can be found in~\cref{app:spatial_discretization_error}. 

\begin{lem}\label{lem:FRDE_potential_spatial_error}
    Let $u$ be the exact solution of~\cref{eqn:FRDE_linear} and $\vec{u}$ be the solution of~\cref{eqn:FRDE_linear_ODE_shifted}. 
    Suppose $u(t,x)$ is $p$-th order spatially continuously differentialable where $p \geq d+\alpha+2$, then 
    \begin{equation}
        \|(u(T,n/N))_{n\in[N]^d} - \vec{u}(T)\| \leq T \frac{2^{p+1} d^{\alpha/2} \max_{t,j}\|\partial_{x_j}^p u(t,x) \|_{L^1}}{\pi^{p-\alpha} N^{p-d-\alpha}}. 
    \end{equation}
\end{lem}

\cref{lem:FRDE_potential_spatial_error} tells that, similar to the case without potential, the order of the spatial discretization error convergence depends on the smoothness of the solution. 
In particular, it can be exponential convergence if the solution is within the Gevrey class. 
The Gevrey class includes infinitely differentiable functions whose $p$-th order derivative grows polynomially in $p!$. 
Notice that Gevrey class greatly enlarges the class of real analytic functions, since the Taylor series will only have convergence radius $0$ if the $p$-th order derivative scales super-linearly in $p!$. 
We give the explicit error bounds and the choice of the grid for bounded errors in the next two results. 
Their proof can be found in~\cref{app:spatial_discretization_error} as well.

\begin{cor}\label{cor:FRDE_potential_spatial_error_exp}
    Consider solving~\cref{eqn:FRDE_linear} on discrete grid points $(T,n/N)$ where $n \in [N]^d$. 
    Suppose that the exact solution $u(t,x)$ is in the Gevrey class $G^{\sigma}$ in the sense that $u(t,x)$ is smooth and there exist constants $\Lambda > 0$ and $\sigma \geq 0$ such that 
    \begin{equation}\label{eqn:gevrey_class}
        \sup_{j\in[d],t\in[0,T],x\in[0,1]^d}|\partial_{x_j}^p u(t,x)| \leq \Lambda^{p+1} (p!)^{\sigma}. 
    \end{equation}
    Then, for any $N \geq (2\Lambda / \pi)(d + \alpha + 2)^{\sigma} $, we have 
    \begin{equation}
        \|(u(T,n/N))_{n\in[N]^d} - \vec{u}(T)\| \leq c_1 T (c_2d)^{c_3 d} d^{\alpha/2} \exp\left( -c_4 N^{1/\sigma} \right),  
    \end{equation}
    where $c_j$'s are constants only depending on $\sigma,\alpha$ and $\Lambda$.  
\end{cor}

\begin{cor}\label{cor:FRDE_potential_spatial_N}
    Consider solving~\cref{eqn:FRDE_linear} on discrete grid points $(T,n/N)$ where $n \in [N]^d$. 
    Suppose that the exact solution $u(t,x)$ is in the Gevrey class $G^{\sigma}$ as in~\cref{cor:FRDE_potential_spatial_error_exp}. 
    Then, in order to bound the spatial discretization error in the quantum state (\emph{i.e.}, $\|\ket{(u(T,n/N))_{n\in[N]^d}} - \ket{\vec{u}(T)}\|$) by $\epsilon$, it suffices to choose 
    \begin{equation}
        N = \mathcal{O} \left( \left( d\log d +  \log\left(\frac{T}{\|(u(T,n/N))_{n\in[N]^d}\|}\right) + \log\left(\frac{1}{\epsilon}\right) \right)^{\sigma} \right). 
    \end{equation}
\end{cor}

Before we proceed to the algorithm design and analysis, we would like to further discuss the relation between the fractional Laplacian $(-\Delta)^{\alpha/2}$ and its discretized version $B$. 
First, notice that by the definition of $B$ in~\cref{eqn:def_B} and~\cref{eqn:def_D}, the oeprator norm of $B$ can be bounded as 
\begin{align}
    \|B\| &= \norm{(\mathcal{F})^{\otimes d}  D (\mathcal{F}^{-1})^{\otimes d} } = \norm{D} = \max_{n\in [N]^d} 2^{\alpha}\pi^{\alpha} \norm{i(n)}^{\alpha} = \max_{n\in[N]^d} 2^{\alpha}\pi^{\alpha} \left( \sum_{j=0}^{d-1} |i_j(n)|^2 \right)^{\alpha/2} \\
    & \leq 2^{\alpha}\pi^{\alpha} \left( \sum_{j=0}^{d-1} |N/2|^2 \right)^{\alpha/2} = \mathcal{O}( d^{\alpha/2} N^{\alpha} ), 
\end{align}
where the inequality in the last line follows from the definition of $i(n)$ in~\cref{eqn:def_i_j}. 
Therefore, although the discretized Laplacian $B$ has a finite operator norm, it grows as the number of grid points along each spatial dimension increases, and tends to infinity when $N \rightarrow \infty$ (in this case $B$ also converges to the continuous fractional Laplacian $(-\Delta)^{\alpha/2}$). 

Fortunately, as shown in~\cref{lem:FRDE_potential_spatial_error},~\cref{cor:FRDE_potential_spatial_error_exp} and~\cref{cor:FRDE_potential_spatial_N}, we can choose a specific value of $N$ to control the discretization error by a target $\epsilon$.  
So, the way in which we resolve the unbounded operator issue is as follows: we first choose a sufficiently large $N$ to ensure that the spatial discretization error is within the error tolerance, and then simply design quantum algorithms for the spatially discretized equation in~\cref{eqn:FRDE_linear_ODE_shifted}. 
The output of quantum algorithms would also be a good approximation of the solution of the original equation in~\cref{eqn:FRDE_linear_shifted}. 
Nevertheless, since $\|B\|$ is larger as we require more accuracy, we need to carefully analyze how the complexities of the quantum algorithms might be affected by $\|B\|$. 
This will be carefully discussed in the following subsections.

\subsection{Second-order Trotter formula}

We solve~\cref{eqn:FRDE_linear_ODE_shifted} by Trotter formula. 
In particular, we divide $[0,T]$ into $r$ equi-length segments and let $h = T/r$. 
Consider 
\begin{equation}
    \mathcal{T} e^{\int_0^T (-B-C(t))dt } \approx \prod_{j=0}^{r-1} S_2((j+1)h,jh)
\end{equation}
where $S_2(jh,(j+1)h)$ is the second-order time-dependent Trotter method, aiming at approximating $\mathcal{T} e^{\int_{jh}^{(j+1)h} (-B-C(t))dt }$ and defined as 
\begin{equation}\label{eqn:2nd_Trotter_def}
    S_2((j+1)h,jh) = e^{-B h/2} e^{-C((j+1/2)h) h}  e^{-B h/2}. 
\end{equation}
In this subsection, we first derive a bound of the Trotter error and show the choice of the number of the segments for bounded errors. 
Then we discuss how to quantumly implement the numerical scheme and estimate its complexity. 

\subsubsection{Error bound, commutator scalings, and the number of the time steps}

We first bound the distance between the exact evolution operator and the numerical integrators. 

\begin{lem}\label{lem:FRDE_linear_potential_Trotter_error}
    Consider solving~\cref{eqn:FRDE_linear_ODE_shifted} using second-order Trotter formula $\prod_{j=0}^{r-1} S_2((j+1)h,jh)$ with time step size $h = T/r$, where the local integrator $S_2$ is defined in~\cref{eqn:2nd_Trotter_def}. 
    Then 
    \begin{equation}\label{eqn:Trotter_error_final_bound}
        \begin{split}
            & \quad \left\|\mathcal{T} e^{\int_0^T (-B-C(t))dt } - \prod_{j=0}^{r-1} S_2((j+1)h,jh)\right\| \\
            & \leq T h^2 \left(  \frac{1}{24}\max\|C''\| +  \frac{1}{4}(\|B\|+\max\|C\|)\max\|C'\|  \right. \\
            & \quad\quad\quad\quad\quad \left. + \frac{1}{6} \max\|[B,[B,C]]\| + \frac{1}{4} \max\|[B,{C}]\|\max\|{C}\| + \frac{1}{3} \max\|{C}\|^3 \right), 
        \end{split}
    \end{equation}
    where all the maximums are taken over $t \in [0,T]$. 
\end{lem}

The proof of~\cref{lem:FRDE_linear_potential_Trotter_error} can be found in~\cref{app:Trotter_error}, which contains two parts. 
First, we deal with the time-ordering operator by bounding the distance between $\mathcal{T} e^{\int_{0}^{h} (-B-C(t))dt }$ and $e^{(-B-C(h/2))h}$ using the variation of parameters formula. 
This part contributes to the first two terms in the error bound (the second line of~\cref{eqn:Trotter_error_final_bound}) involving time derivatives of the potential matrix $C(t)$. 
Then we can bound the error between $e^{(-B-C(h/2))h}$ and $S_2$ by time-independent Trotter error bounds. 
Notice that the proof for the second part is different from that for Hamiltonian simulation~\cite{ChildsSuTranEtAl2020}, since in our case the parameters $\beta$ and $\gamma$ in the exponentials $e^{-\beta B}$ and $e^{-\gamma C}$ are restricted to non-negative to avoid exponential overhead, while for Hamiltonian simulation there is no such restriction. 
Instead, we mostly follow the procedure in~\cite{JahnkeLubich2000} to establish the second part of the error bound. 
This contributes to the last three terms in the error bound (the third line of~\cref{eqn:Trotter_error_final_bound}) involving the commutators and the spectral norm of $C$ explicitly.

Now we compute the norm of the commutators $[B, {C}(t)]$ and $[B, [B,{C}(t)]]$. 
All the results are for a fixed $t$, so we will omit the explicit $t$ dependence in our notation for now. 
Naive bounds are $\|[B,{C}]\| \leq \mathcal{O}(\|B\|) = \mathcal{O}(d^{\alpha/2}N^{\alpha})$ and $\|[B, [B,{C}]]\| \leq \mathcal{O}(\|B\|^2) = \mathcal{O}(d^{\alpha}N^{2\alpha})$. 
However, these naive bounds are improvable, because the order of the commutator of the Laplacian operator can be reduced. 
Such a phenomenon has been observed in~\cite{JahnkeLubich2000,AnFangLin2021}. 
To see this, let us take $\alpha = 2$ and consider its continuous analog. 
Then for any smooth function $f(x)$, 
\begin{equation}
\begin{split}
    [-\Delta, {c}] f &= -\Delta ({c} f) - {c} (-\Delta f) \\
    & = -(\Delta {c}) f - 2(\nabla {c}) \cdot (\nabla f) - {c} \Delta f + {c} \Delta f \\
    & = -(\Delta {c}) f - 2(\nabla {c}) \cdot (\nabla f). 
\end{split}
\end{equation}
So we expect, in the discrete setting, the first commutator is bounded by the discretized divergence operator, whose norm is only $\mathcal{O}(dN)$. 

Now we state the improved bound in the discrete setting. 
The proof can be found in~\cref{app:commutator}, which is a generalization of~\cite{JahnkeLubich2000}.

\begin{lem}\label{lem:commutator}
    Suppose that ${c}(t,x)$ is a bounded $C^{5+d}$ function in $x$. 
    Then we have 
    \begin{equation}
        \|[B,{C}(t)]\| \leq \mathcal{O}(d^{\alpha/2}N^{\alpha/2})
    \end{equation}
    and 
    \begin{equation}
        \|[B, [B,{C}(t)]]\| \leq \mathcal{O}(d^{\alpha}N^{\alpha}). 
    \end{equation}
\end{lem}

The choice of the number of time steps is a direct consequence of~\cref{lem:FRDE_linear_potential_Trotter_error} and~\cref{lem:commutator}. 

\begin{cor}\label{cor:FRDE_linear_potential_r}
    Consider solving~\cref{eqn:FRDE_linear_ODE_shifted} using second-order Trotter formula $\prod_{j=0}^{r-1} S_2((j+1)h,jh)$ with time step size $h = T/r$, where the local integrator $S_2$ is defined in~\cref{eqn:2nd_Trotter_def}. 
    Suppose that $\|{C}^{(k)}(t)\|$'s are uniformly bounded in $t$ for $k \leq d+5$. 
    Then, in order to bound the operator splitting error by $\epsilon$, it suffices to choose 
    \begin{equation}
        r = \mathcal{O}\left( d^{\alpha/2} N^{\alpha/2} \frac{T^{3/2}}{\epsilon^{1/2}}  \right). 
    \end{equation} 
\end{cor}

\subsubsection{Quantum implementation and complexity estimate}

Now we discuss quantum implementation of this method. 
The main idea is to construct the block-encodings of $e^{-B h/2}$ and $e^{-{C}((j+1/2)h) h}$ and multiply them together using~\cref{lem:block_encoding_product}. 
The block-encodings can be constructed using controlled rotations since each evolution operator is unitarily equivalent to a diagonal matrix and the corresponding unitary transformation matrix is efficiently implementable. 
The nuance is that the operator $e^{-{C}((j+1/2)h) h}$ still depends on the specific time, so we use the counter register in the compression gadget as the time clock as well. 

Suppose that we are given the oracle $O_1$, defined in~\cref{eqn:def_O_high_d_linear}, that encodes the eigenvalues of $B$, and the oracle $O_{2}$ that gives the element of $C(t)$ as 
\begin{equation}\label{eqn:def_O_high_d_potential}
    O_2: \ket{t}\ket{n_0}\cdots \ket{n_{d-1}}\ket{0} \rightarrow \ket{t}\ket{n_0}\cdots \ket{n_{d-1}}\ket{c(t,n/N)}. 
\end{equation}
Here, with an abuse of notations, $\ket{t}$ represents some specific way of encoding the information of $t$, \emph{i.e.}, $(2j+1)$ for the time point $(j+1/2)h$ after time discretization (the reason why it is $(2j+1)$ will be clear soon). 

As discussed before, we may use the circuit in~\cref{fig:circuit_FRDE_linear} with a replacement of $O_{\exp,1}$ by $O_{\exp,2}: \ket{x}\ket{0} \rightarrow \ket{x}\ket{e^{-xh/2}}$ to construct a $(1,*,0)$-block-encoding of $e^{-Bh/2}$. 
Here the number of the ancilla qubits depends on those for encoding the eigenvalues of $B$ and their exponentials, on which we do not keep track for technical simplicity. 
We denote this block-encoding by $V_{B,h/2}$. 
Similarly, a $(1,*,0)$-block-encoding of $e^{-Bh}$ can be constructed with $O_{\exp,3}: \ket{x}\ket{0} \rightarrow \ket{x}\ket{e^{-xh}}$. 
We denote it by $V_{B,h}$. 
Furthermore, in~\cref{fig:circuit_FRDE_linear}, if we discard the QFT steps, and replace $O_{\exp,1}$ by $O_{\exp,3}$ and the oracle $O_1$ by $O_2$ controlled by an extra counter register, we can construct a (controlled version of) $(1,*,0)$-block-encoding of $e^{-C((j+1/2)h) h}$, denoted by $V_{C,j}$. 
Notice that the constructions of block-encodings $V_{B,h/2}$, $V_{B,h}$ and $V_{C,j}$ only require $\mathcal{O}(1)$ queries to the aforementioned oracles and QFT. 

\begin{figure}
    \centerline{
    
    \Qcircuit @R=1em @C=1em {
    \text{Counter} \quad\quad\quad\quad\quad & \qw & \gate{\text{ADD}} & \ctrl{1} & \gate{\text{ADD}} & \qw & \gate{\text{ADD}} & \ctrl{1} & \gate{\text{ADD}} &\qw & \cdots \\
    \text{Ancilla} \quad\quad\quad\quad\quad & \multigate{1}{V_{B,h/2}} & \ctrlo{-1} & \multigate{1}{V_{C,0}} & \ctrlo{-1} & \multigate{1}{V_{B,h}} & \ctrlo{-1} & \multigate{1}{V_{C,1}} & \ctrlo{-1} & \qw & \cdots\\ 
    \text{System}  \quad\quad\quad\quad\quad & \ghost{V_{B,h/2}} & \qw & \ghost{{V_{C,0}}} & \qw & \ghost{V_{B,h}} & \qw & \ghost{V_{C,1}} & \qw & \qw & \cdots \\
     \\
    \text{Counter} \quad\quad\quad\quad\quad & \cdots\quad\quad & \qw & \gate{\text{ADD}} & \ctrl{1} & \gate{\text{ADD}} & \qw & \gate{\text{ADD}} & \gate{(\text{ADD}^{\dagger})^{2r+1}} & \qw \\
    \text{Ancilla} \quad\quad\quad\quad\quad & \cdots\quad\quad & \multigate{1}{V_{B,h}} & \ctrlo{-1} & \multigate{1}{V_{C,{j-1}}} & \ctrlo{-1} & \multigate{1}{V_{B,h/2}} & \ctrlo{-1} & \qw & \qw \\
    \text{System}  \quad\quad\quad\quad\quad & \cdots\quad\quad & \ghost{V_{B,h}} & \qw & \ghost{V_{C,{j-1}}} & \qw & \ghost{{V_{B,h/2}}} & \qw & \qw & \qw 
    }
    }
    \caption{ Quantum circuit for implementing second-order Trotter method. Here the Counter register contains $\lceil\log_2 (2r+1) \rceil + 1$ qubits. ADD implements addition by $1$ modulo $2^{\lceil\log_2 (2r+1) \rceil + 1}$. }
    \label{fig:2nd_Trotter}
\end{figure}
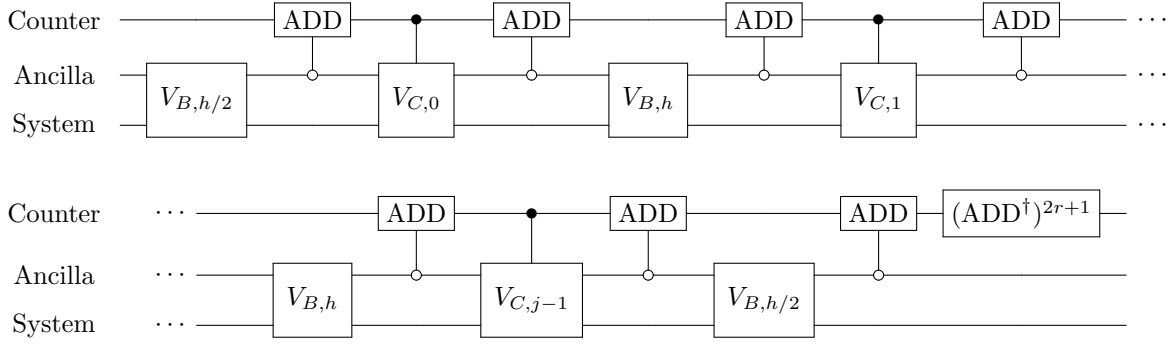

Now we construct the numerical integrator. 
Mathematically we need to block encode the operator 
\begin{equation}
    \prod_{j=0}^{r-1} S_2((j+1)h,jh) = e^{-Bh/2}  \left(\prod_{j=1}^{r-1} e^{-C((j+1/2)h)h }  e^{-Bh} \right) e^{-C(h/2) h } e^{-Bh/2}. 
\end{equation}
This can be done by the circuit in~\cref{fig:2nd_Trotter}, which can be viewed as a variant of the compression gadget in~\cref{lem:block_encoding_product}. 
The idea is to use an extra counter register with $\lceil\log_2 (2r+1) \rceil + 1$ qubits for both keeping track of the success/failure of the multiplication and indicating the index of the time step. 
If all the applications of the block-encodings of the local exponentials are successful, then the value of the counter register at each step suggests the correct time step index, and the final value is reset to be $0$. 
Notice that the block-encoding factor of $V_{B,h}$ and $V_{C,j}$ are $1$. 
Then the circuit in~\cref{fig:2nd_Trotter} gives a $(1,*,0)$-block-encoding of $\prod_{j=0}^{r-1} S_2((j+1)h,jh)$, as desired.

We apply this block-encoding to the input state $\ket{u_0}$, and the final state is 
\begin{equation}\label{eqn:Trotter_overall_final}
    \frac{1}{\|\vec{u}_0\|} \sum_{n\in[N]^d}  ( \vec{u}_{r} )_n \ket{n_0}\cdots\ket{n_{d-1}} \ket{0} + \ket{\perp}, 
\end{equation}
where 
\begin{equation}
\begin{split}
    \vec{u}_r = \prod_{j=0}^{r-1} S_2((j+1)h,jh) \vec{u}_0 \approx \vec{u}(T), 
\end{split}
\end{equation}
and $\ket{\perp}$ represents the junk state with ancilla register not equal to $0$. 
The final step is to measure the ancilla registers of~\cref{eqn:Trotter_overall_final}. 
If all the ancilla registers are $0$, then we get a good approximation of $\ket{u(T)}$. 
The averaged number of repeats after amplitude amplification is $\mathcal{O}(\|\vec{u}_0\|/\|\vec{u}(T)\|)$.

We summarize the overall complexity as follows.
\begin{thm}\label{thm:trotter_complexity}
    Consider solving~\cref{eqn:FRDE_linear} on discrete grid points $(T,n/N)$ where $n \in [N]^d$. 
    Let $u(t,x)$ denote the solution of the equation~\cref{eqn:FRDE_linear}, and $\vec{u}(t)$ denote the solution of the spatially discretized~\cref{eqn:FRDE_linear_ODE_shifted}. 
    Suppose that 
    \begin{enumerate}
        \item we are given oracles $O_1$ and $O_2$ defined in~\cref{eqn:def_O_high_d_linear} and~\cref{eqn:def_O_high_d_potential}, and the state preparation oracle $O_u: \ket{0} \rightarrow \ket{\vec{u}_0}$, 
        \item $u(t,x)$ is in the Gevrey class $G^{\sigma}$ in the sense that $u(t,x)$ is smooth and there exist constants $\Lambda > 0$ and $\sigma \geq 0$ such that 
    \begin{equation}
        \sup_{j\in[d],t\in[0,T],x\in[0,1]^d}|\partial_{x_j}^p u(t,x)| \leq \Lambda^{p+1} (p!)^{\sigma}. 
    \end{equation}
    \item $\|(u(T,n/N))_{n\in[N]^d}\| \geq \widetilde{g}(T)$ for a function $\widetilde{g}$. 
    \item $\|\vec{u}(0)\|/\|\vec{u}(T)\| \leq g(T)$ for a function $g$. 
    \end{enumerate}
    Then, with second-order operator splitting method for time propagation, an $\epsilon$-approximation of $\ket{(u(T,n/N))_{n\in[N]^d}}$ can be obtained by choosing 
    \begin{equation}
        N = \mathcal{O} \left( \left( d\log d +  \log\left(\frac{T}{\widetilde{g}(T)}\right) + \log\left(\frac{1}{\epsilon}\right) \right)^{\sigma} \right).
    \end{equation}
    and using 
    \begin{enumerate}
        \item 
        \begin{equation}
            \mathcal{O}\left((g(T))^{3/2} \frac{T^{3/2}}{\epsilon^{1/2}} d^{\alpha/2} \left( d\log d +  \log\left(\frac{T}{\widetilde{g}(T)}\right) + \log\left(\frac{1}{\epsilon}\right) \right)^{\alpha\sigma/2}  \right)
        \end{equation}
        queries to $O_1$,$O_2$, their inverses and controlled versions, 
        \item 
        \begin{equation}
            \mathcal{O}\left(g(T)\right)
        \end{equation}
        queries to the state preparation oracle $O_u$ and its inverse. 
        \item 
        \begin{equation}
        \begin{split}
            \mathcal{O}\left( (g(T))^{3/2} \frac{T^{3/2}}{\epsilon^{1/2}} d^{\alpha/2+1} \left( d\log d +  \log\left(\frac{T}{\widetilde{g}(T)}\right) + \log\left(\frac{1}{\epsilon}\right) \right)^{\alpha\sigma/2} \right. \\
            \left. \times \log^2\left( d\log d +  \log\left(\frac{T}{\widetilde{g}(T)}\right) + \log\left(\frac{1}{\epsilon}\right) \right) \right)
        \end{split}
        \end{equation}
        additional elementary gates. 
    \end{enumerate}
\end{thm}
\begin{proof}
    It suffices to bound both spatial and time discretization errors by $\epsilon$. 
    The choice of $N$ directly follows~\cref{cor:FRDE_potential_spatial_N}. 
    We now count the overall complexity. 
    Each run of the algorithm requires an application of~\cref{fig:circuit_FRDE_linear} which implements $e^{-Bh/2}$ and $e^{-{C}(t) h}$ for $\mathcal{O}(r)$ times. 
    In each block-encoding of $e^{-Bh/2}$ and $e^{-{C}(t) h}$, we need to use $\mathcal{O}(1)$ queries to $O_1$ and $O_2$, and $\mathcal{O}(d\log^2(N))$ additional gates mainly due to the QFT. 
    According to~\cref{cor:FRDE_linear_potential_r}, it suffices to choose $r = \mathcal{O}\left( g(T)^{1/2} d^{\alpha/2} N^{\alpha/2} \frac{T^{3/2}}{\epsilon^{1/2}}  \right)$. 
    Here the extra factor $g(T)^{1/2}$ is because~\cref{cor:FRDE_linear_potential_r} only bounds the operator norm, and in order to bound the error in the quantum state by $\epsilon$, the operator norm error bound needs to be bounded by $\mathcal{O}(\epsilon \|\vec{u}(T)\|/\|\vec{u}(0)\|)$ according to~\cref{lem:error_vector_quantum_state}.  
    The averaged number of repeats to succeed after amplitude amplification is $\mathcal{O}(\|\vec{u}(0)\|/\|\vec{u}(T)\|) = \mathcal{O}(g(T))$. 
    Multiplying these together gives the overall query complexity and additional gates required. 
    Notice that in each run we only need one query to the state preparation oracle, so the overall number of state preparation is only $\sim g(T)$. 
\end{proof}

\subsection{Time-marching method}

Now we consider an alternative algorithm proposed in~\cite{FangLinTong2022} called time-marching method. 
The method is designed for general ODE 
\begin{equation}\label{eqn:ODE_general_At}
    \frac{d}{dt} \phi(t) = A(t) \phi(t), \quad \phi(0) = \phi_0
\end{equation}
with a time-dependent matrix-valued function $A(t)$, and is a quantum analog of the classical exponential propagation methods. 
It first divides the time interval $[0,T]$ into small segments with mesh $0 = t_0 < t_1 < \cdots < t_L = T$ and applies the short-time evolution operator sequentially. 
While naive applications of a sequence of non-unitary operators may incur an exponential overhead in the number of the operators, the time-marching method avoids such overhead by a technique that combines the uniform singular value transformation and the amplitude amplification. 
We refer to~\cite{FangLinTong2022} for more details and only roughly summarize the main result here. 
\begin{lem}[Theorem 8 of~\cite{FangLinTong2022}]\label{lem:time_marching}
    Consider the ODE~\cref{eqn:ODE_general_At}. 
    Suppose that we are given the prepare oracle $O_{\phi}$ of $\ket{\phi_0}$ such that $O_{\phi} \ket{0} = \ket{\phi_0}$ and an input model of $A(t)$, denoted by $\text{MAT}_A$, that simultaneously block encodes $A(t_k')$ at some refined mesh points $t_k'$. 
    Then, with the time-marching method, an $\epsilon$-approximation of $\ket{\phi(T)}$ can be prepared using 
    \begin{equation}
        \mathcal{O}\left(\eta^2 T^2 Q \log(\eta T Q) \frac{\log(\eta T Q/\epsilon)}{\log\log(\eta T Q/\epsilon)} \right)
    \end{equation}
    queries to $\text{MAT}_A$ and $\mathcal{O}(Q)$ queries to $O_{\phi}$, its inverse and controlled version. 
    Here $\eta$ is the block-encoding factor of $\text{MAT}_A$ such that $\eta \geq \|A(t)\|$ for all $t\in[0,T]$, and 
    \begin{equation}
        Q = \frac{\|\phi(0)\|\prod_{l=1}^L \left\|\mathcal{T} e^{\int_{t_{l-1}}^{t_l} A(t)dt }\right\|}{\|\phi(T)\|}. 
    \end{equation}
\end{lem}

Now we discuss the complexity of applying the time-marching method to spatially discretized equation~\cref{eqn:FRDE_linear_ODE_shifted}. 
For technical simplicity, we only estimate the query complexities. 

\begin{thm}\label{thm:time_marching_FRDE}
    Consider solving~\cref{eqn:FRDE_linear} on discrete grid points $(T,n/N)$ where $n \in [N]^d$. 
    Let $u(t,x)$ denote the solution of the equation~\cref{eqn:FRDE_linear}, and $\vec{u}(t)$ denote the solution of the spatially discretized~\cref{eqn:FRDE_linear_ODE_shifted}. 
    Suppose that 
    \begin{enumerate}
        \item we are given oracles encoding the eigenvalues of $B$ and the diagonal entries of ${C}(t)$, \emph{i.e.},  $O_1$ and $O_2$ defined in~\cref{eqn:def_O_high_d_linear} and~\cref{eqn:def_O_high_d_potential}, and the state preparation oracle $O_u: \ket{0} \rightarrow \ket{\vec{u}_0}$, 
        \item $v(t,x)$ is in the Gevrey class $G^{\sigma}$ in the sense that $v(t,x)$ is smooth and there exist constants $\Lambda > 0$ and $\sigma \geq 0$ such that 
    \begin{equation}
        \sup_{j\in[d],t\in[0,T],x\in[0,1]^d}|\partial_{x_j}^p v(t,x)| \leq \Lambda^{p+1} (p!)^{\sigma}. 
    \end{equation}
    \item $\|(v(T,n/N))_{n\in[N]^d}\| \geq \widetilde{g}(T)$ for a function $\widetilde{g}$. 
    \end{enumerate}
    Then, with the time-marching method for time propagation, an $\epsilon$-approximation of $\ket{(u(T,n/N))_{n\in[N]^d}}$ can be obtained by choosing 
    \begin{equation}
        N = \mathcal{O} \left( \left( d\log d +  \log\left(\frac{T}{\widetilde{g}(T)}\right) + \log\left(\frac{1}{\epsilon}\right) \right)^{\sigma} \right).
    \end{equation}
    and using 
    \begin{enumerate}
        \item 
        \begin{equation}
            \widetilde{\mathcal{O}}\left( Q T^2 d^{\alpha} \left( d\log d +  \log\left(\frac{T}{\widetilde{g}(T)}\right) + \log\left(\frac{1}{\epsilon}\right) \right)^{2\alpha\sigma} \log\left(\frac{1}{\epsilon}\right) \right)
        \end{equation}
        queries to ${O}_1$, ${O}_2$, the $d$-dimensional QFT circuit, their inverses and controlled versions, where 
        \begin{equation}\label{eqn:time_marching_Q_def}
        Q = \frac{\|\vec{u}(0)\|\prod_{l=1}^L \left\|\mathcal{T} e^{\int_{t_{l-1}}^{t_l} (-B-{C}(t))dt }\right\|}{\|\vec{u}(T)\|},  
        \end{equation}
        \item 
        \begin{equation}
            \mathcal{O}\left(Q\right)
        \end{equation}
        queries to the state preparation oracle $O_u$ or its inverse. 
    \end{enumerate}
\end{thm}
\begin{proof}
    First, the oracle $\text{MAT}_A$ mentioned in~\cref{lem:time_marching} can be constructed through the given ${O}_1$ and ${O}_2$ and additional $\mathcal{O}(d\log^2(N))$ gates. 
    The idea is as follows. 
    We can first construct the block-encoding of the diagonal matrix $D$ with ${O}_1$, then construct the block-encoding of $B = (\mathcal{F})^{\otimes d} D (\mathcal{F}^{-1})^{\otimes d}$ using QFT. 
    Meanwhile we may construct the simultaneous block-encoding of ${C}(t)$ with ${O}_2$. 
    Finally, we can linearly combine these two block-encodings to obtain the desired $\text{MAT}_A$ that block encodes $A(t) = -B - {C}(t)$. 
    According to~\cite[Lemma 48 \& Lemma 52 \& Lemma53]{GilyenSuLowEtAl2019}, such approach requires $\mathcal{O}(1)$ queries to ${O}_1$ and ${O}_2$, so we may directly estimate the number of queries to $\text{MAT}_A$ as that of queries to ${O}_1$ and ${O}_2$. 
    Note that there are additional $\mathcal{O}(d\log^2(N))$ gates required to construct $\text{MAT}_A$ from ${O}_1$ and ${O}_2$, but we only focus on the query complexities here for technical simplicity. 
    
    Now we may directly use~\cref{lem:time_marching} to estimate the complexity, and it suffices to write down explicit scalings of the block-encoding factor $\eta$ in the example of~\cref{eqn:FRDE_linear_ODE_shifted}. 
    Under the assumption that ${C}(t)$ is uniformly bounded, the spectral norm $\|-B-{C}(t)\| = \mathcal{O}(d^{\alpha/2}N^{\alpha})$, so the parameter $\eta = \mathcal{O}(d^{\alpha/2}N^{\alpha})$. 
    Using the choice of $N$ estimated in~\cref{lem:FRDE_potential_spatial_error}, we have 
    \begin{equation}
        \eta = \mathcal{O}\left( d^{\alpha/2} \left( d\log d +  \log\left(\frac{T}{\widetilde{g}(T)}\right) + \log\left(\frac{1}{\epsilon}\right) \right)^{\alpha\sigma} \right). 
    \end{equation}
    Plugging this parameter back to the scalings in~\cref{lem:time_marching} yields the desired estimates. 
\end{proof}

\subsection{Truncated Dyson series method}

Now we discuss the query complexity of applying the state-of-the-art generic quantum ODE solvers to our linear fractional reaction-diffusion equation. 
We consider the method proposed in~\cite{BerryCosta2022}, which is based on the truncated Dyson series. 
It first expands the solution via truncated Dyson series and encode it into a linear system of equations, then solve it using the optimal quantum linear system algorithm. 
It works for the most general linear ODE with time dependent coefficient matrix and possible inhomogeneous term. 
In our case, we are only interested in the homogeneous equation, so here we only summarize their main result for the homogeneous equation~\cref{eqn:ODE_general_At}. 

\begin{lem}[Theorem 1 of~\cite{BerryCosta2022}]\label{lem:Dyson_method}
    Consider the ODE~\cref{eqn:ODE_general_At} where the coefficient matrix $A(t)$ has non-positive logarithmic norm for all $t$.  
    Suppose that we are given the prepare oracle $O_{\phi}$ of $\ket{\phi_0}$ such that $O_{\phi} \ket{0} = \ket{\phi_0}$ and an input model of $A(t)$, denoted by $\text{MAT}_A$, that simultaneously block-encoding $A(t_k')$ at some refined mesh points $t_k'$. 
    Then, with the truncated Dyson series method, an $\epsilon$-approximation of $\ket{\phi(T)}$ can be prepared using 
    \begin{equation}
        \mathcal{O}\left( \frac{\max_{t\in[0,T]} \|\phi(t)\|}{\|\phi(T)\|} \eta T \log(1/\epsilon) \log(\eta T/\epsilon) \right)
    \end{equation}
    queries to $\text{MAT}_A$ and 
    \begin{equation}
        \mathcal{O}\left( \frac{\max_{t\in[0,T]} \|\phi(t)\|}{\|\phi(T)\|} \eta T \log(1/\epsilon) \right)
    \end{equation}
    queries to $O_{\phi}$. 
    Here $\eta$ is the block-encoding factor of $\text{MAT}_A$ such that $\eta \geq \|A(t)\|$ for all $t\in[0,T]$. 
\end{lem}

\begin{thm}\label{thm:Dyson_method_FRDE}
    Consider solving~\cref{eqn:FRDE_linear} on discrete grid points $(T,n/N)$ where $n \in [N]^d$. 
    Let $u(t,x)$ denote the solution of the equation~\cref{eqn:FRDE_linear}, and $\vec{u}(t)$ denote the solution of the spatially discretized~\cref{eqn:FRDE_linear_ODE_shifted}. 
    Suppose that 
    \begin{enumerate}
        \item we are given oracles encoding the eigenvalues of $B$ and the diagonal entries of ${C}(t)$, \emph{i.e.},  $O_1$ and $O_2$ defined in~\cref{eqn:def_O_high_d_linear} and~\cref{eqn:def_O_high_d_potential}, and the state preparation oracle $O_u: \ket{0} \rightarrow \ket{\vec{u}_0}$, 
        \item $u(t,x)$ is in the Gevrey class $G^{\sigma}$ in the sense that $u(t,x)$ is smooth and there exist constants $\Lambda > 0$ and $\sigma \geq 0$ such that 
    \begin{equation}
        \sup_{j\in[d],t\in[0,T],x\in[0,1]^d}|\partial_{x_j}^p u(t,x)| \leq \Lambda^{p+1} (p!)^{\sigma}. 
    \end{equation}
    \item $\|(u(T,n/N))_{n\in[N]^d}\| \geq \widetilde{g}(T)$ for a function $\widetilde{g}$,  
     \item $\|\vec{u}(0)\|/\|\vec{u}(T)\| \leq g(T)$ for a function $g$. 
    \end{enumerate}
    Then, with the truncated Dyson series method for time propagation, an $\epsilon$-approximation of \\ $\ket{(u(T,n/N))_{n\in[N]^d}}$ can be obtained by choosing 
    \begin{equation}
        N = \mathcal{O} \left( \left( d\log d +  \log\left(\frac{T}{\widetilde{g}(T)}\right) + \log\left(\frac{1}{\epsilon}\right) \right)^{\sigma} \right).
    \end{equation}
    and using 
    \begin{enumerate}
        \item 
        \begin{equation}
            \widetilde{\mathcal{O}}\left( g(T) T  d^{\alpha/2} \left( d\log d +  \log\left(\frac{T}{\widetilde{g}(T)}\right) + \log\left(\frac{1}{\epsilon}\right) \right)^{\alpha\sigma} \left(\log\left(\frac{1}{\epsilon}\right)\right)^2  \right)
        \end{equation}
        queries to ${O}_1$ and ${O}_2$, $d$-dimensional QFT,  their inverses and controlled versions, 
        \item 
        \begin{equation}
            \mathcal{O}\left( g(T) T  d^{\alpha/2} \left( d\log d +  \log\left(\frac{T}{\widetilde{g}(T)}\right) + \log\left(\frac{1}{\epsilon}\right) \right)^{\alpha\sigma} \log\left(\frac{1}{\epsilon}\right)  \right)
        \end{equation}
        queries to the state preparation oracle $O_u$ or its inverse. 
    \end{enumerate}
\end{thm}
\begin{proof}
    As shown in the proof of~\cref{thm:time_marching_FRDE}, we may directly estimate the number of queries to ${O}_j$ as that of queries to $\text{MAT}_A$, and the block-encoding factor $\eta$ can be bounded as 
    \begin{equation}
        \eta = \mathcal{O}\left( d^{\alpha/2} \left( d\log d +  \log\left(\frac{T}{\widetilde{g}(T)}\right) + \log\left(\frac{1}{\epsilon}\right) \right)^{\alpha\sigma} \right). 
    \end{equation}
    Since the coefficient matrix $(-B-{C}(t))$ is always negative semi-definite, the norm of the solution $\|\phi(t)\|$ is non-increasing over $t$, so 
    \begin{equation}
        \frac{\max_{t\in[0,T]} \|\vec{u}(t)\|}{\|\vec{u}(T)\|} = \frac{ \|\vec{u}(0)\|}{\|\vec{u}(T)\|} \leq g(T). 
    \end{equation}
    Plugging these parameters back to~\cref{lem:Dyson_method} completes the proof. 
\end{proof}

\subsection{Linear combination of Hamiltonian simulation in the interaction picture}

We have discussed and analyzed the second-order operator splitting method, the time-marching method, and the truncated Dyson series method. 
All the methods have extra polynomial dependence on the dimension $d$, which comes from the dependence on the spectral norm of the discrete fractional Laplacian operator $B$, although the operator splitting method may partially benefit from its commutator scalings. 

In quantum dynamics, a common technique to avoid the explicit dependence on $\|B\|$ is to simulate the dynamics in the interaction picture. 
Specifically, if we consider the fractional Schr\"odinger equation 
\begin{equation}
    i \frac{d}{dt} \ket{\psi} = (-B - {C}(t)) \ket{\psi}, 
\end{equation}
then, by defining the interaction picture Hamiltonian $H_I(t) = - e^{-iBt} {C}(t) e^{iBt}$ and $\ket{\psi_I} = e^{-iBt}\ket{\psi}$, we may obtain the transformed solution $\ket{\psi_I}$ by simulating $i \frac{d}{dt} \ket{\psi_I} = H_I(t) \ket{\psi_I}$. 
Here $H_I(t)$ is bounded independently of $\|B\|$, and its oscillations depend on $\|B\|$. 
Therefore we may efficiently simulate the interaction picture Hamiltonian using truncated Dyson series method which has linearly dependence on $\|H_I\|$ but only poly-logarithmically depends on its derivatives. 
We refer to~\cite{LowWiebe2019} for more details. 

The success of the interaction picture Hamiltonian simulation relies on the fast-forwarded implementation of $e^{iBs}$, that is, $e^{iBs}$ can be implemented for any real number $s$ with cost independent of $\|B\|$ and $|s|$. 
However, in our reaction-diffusion equation, which can be viewed as the imaginary time evolution of the Schr\"odinger equation, we cannot directly apply similar technique. 
This is because $e^{-Bs}$ is only fast-forwardable when $s \geq 0$, while the transformation into the analog of interaction picture requires both forward and backward time evolution.

\subsubsection{Representation}

To take advantage of the interaction picture technique, we can relate the reaction-diffusion equation with the Hamiltonian simulation problem. 
A recent work~\cite{AnLiuLin2023} shows that any linear ODE can be represented as a linear combination of Hamiltonian simulation. 
In particular, we may write the evolution operator of our fractional reaction-diffusion equation as 
\begin{equation}\label{eqn:LCHS}
    \mathcal{T} e^{ - \int_0^T (B+{C}(s)) ds } = \int_{\mathbb{R}} \frac{1}{\pi(1+\xi^2)} \mathcal{T} e^{-i \int_0^T  \xi(B+{C}(s)) ds } d\xi. 
\end{equation}
Here $\mathcal{T}$ is the time-ordering operator. 
The proof of~\cref{eqn:LCHS} can be found in~\cite{AnLiuLin2023}. 
Now we use the interaction picture simulation. 
Let 
\begin{equation}\label{eqn:LCHS_def_U}
    U(t) =  \mathcal{T} e^{-i \int_0^t  \xi(B+{C}(s)) ds }, 
\end{equation} 
then $U(t)$ satisfies the time-dependent Hamiltonian simulation problem as  
\begin{equation}\label{eqn:LCHS_def_U_ODE}
    \frac{dU}{dt} = -i (\xi B + \xi {C}(t)) U(t), \quad U(0) = I. 
\end{equation}
Let 
\begin{equation}
    U_I(t) = e^{i\xi B t} U(t). 
\end{equation}
We may compute that 
\begin{equation}
    \begin{split}
        \frac{d U_I}{dt} &= i \xi  B e^{i\xi B t} U(t) - i e^{i\xi B t} (\xi B + \xi {C}(t)) U(t) \\
        & = - i e^{i\xi B t} \xi {C}(t) e^{-i\xi B t} U_I(t). 
    \end{split}
\end{equation}
Define 
\begin{equation}\label{eqn:Hamiltonian_IP}
    H_I(t;\xi) = e^{i\xi B t} \xi {C}(t) e^{-i\xi B t}. 
\end{equation}
Then 
\begin{equation}
    U_I(t) = \mathcal{T} e^{-i \int_0^t H_I(s;\xi) ds},  
\end{equation}
and we may write~\cref{eqn:LCHS} as 
\begin{equation}\label{eqn:LCHS_IP}
    \mathcal{T} e^{ - \int_0^T (B+{C}(s)) ds } = \int_{\mathbb{R}} \frac{1}{\pi(1+\xi^2)} e^{-i\xi B T} \mathcal{T} e^{-i \int_0^T H_I(s;\xi) ds} d\xi. 
\end{equation}

\subsubsection{Numerical quadrature}

We can truncate~\cref{eqn:LCHS_IP} over a finite interval $[-\Xi,\Xi]$ and write it as 
\begin{equation}
    \mathcal{T} e^{ - \int_0^T (B+{C}(s)) ds } \approx \int_{-\Xi}^{\Xi} \frac{1}{\pi(1+\xi^2)} e^{-i\xi B T} \mathcal{T} e^{-i \int_0^T H_I(s;\xi) ds} d\xi. 
\end{equation}
The resulting integral can be discretized using standard numerical quadrature. 
Here we use the simplest Riemann sum formula with $M$ grid points. 
For $0 \leq j \leq M-1$, let $\xi_j = -\Xi + 2j \Xi/M$ and $w_j = \frac{1}{\pi (1+\xi_j^2)} \frac{2\Xi}{M}$. 
Then 
\begin{equation}\label{eqn:LCHS_discrete}
    \mathcal{T} e^{ - \int_0^T (B+{C}(s)) ds } \approx \sum_{j=0}^{M-1} w_j e^{-i\xi_j B T} \mathcal{T} e^{-i \int_0^T H_I(s;\xi_j) ds}. 
\end{equation}
The discretization error can be bounded as follows, and its proof is given in~\cref{app:LCHS_discretization_error}. 

\begin{lem}\label{lem:LCHS_discretization_error}
    We have 
    \begin{equation}
        \left\| \mathcal{T} e^{ - \int_0^T (B+{C}(s)) ds } - \sum_{j=0}^{M-1} w_j e^{-i\xi_j B T} \mathcal{T} e^{-i \int_0^T H_I(s;\xi_j) ds} \right\| \leq \frac{2}{\pi \Xi} + \frac{2\Xi^2}{\pi M} \left( 1 + T (\|B\|+\max\|{C}\|) \right). 
    \end{equation}
    In order to bound the discretization error by $\mathcal{O}(\epsilon)$, it suffices to choose 
    \begin{equation}
        \Xi = \mathcal{O}\left( \frac{1}{\epsilon} \right), \quad M = \mathcal{O}\left( \frac{T\|B\|}{\epsilon^3} \right). 
    \end{equation}
\end{lem}

\subsubsection{Implementation and complexity}

Suppose that we are given the same input oracles as in previous algorithms, encoding the eigenvalues of $B$ and the diagonal entries of ${C}(t)$, \emph{i.e.},  $O_1$ and $O_2$ defined in~\cref{eqn:def_O_high_d_linear} and~\cref{eqn:def_O_high_d_potential}. 
The sketch of the algorithm is to simulate the interaction picture Hamiltonian using truncated Dyson series method~\cite{LowWiebe2019} and then apply the LCU technique to compute~\cref{eqn:LCHS_discrete}. 

We start with the HAM-T encoding of the matrix $H_I$. 
For a fixed time step size $h$ and an integer such that $[mh,(m+1)h] \subset [0,T]$, we may first construct the HAM-T encoding of ${C}$ from its sparse input oracle following the approach in~\cite{GilyenSuLowEtAl2019}. 
The resulting HAM-T encoding satisfies 
\begin{equation}
    \bra{0}_a \text{HAM-T}_{C,m} \ket{0}_a = \sum_{l=0}^{M_H-1} \ket{l}\bra{l} \otimes \frac{{C}(mh+lh/M_H)}{\alpha_C}. 
\end{equation}
Here $\alpha_C$ is the upper bound of $\|{C}\|$, and $M_H$ is the number of grid points used in the truncated Dyson series method. 
By appending an ancilla register $\ket{j}$ encoding the index for $\xi$ and applying the controlled rotation $\ket{j}\ket{0} \rightarrow \ket{j} \left(\frac{\xi_j}{\Xi}\ket{0} + \sqrt{1-\frac{\xi_j^2}{\Xi^2}} \ket{1}\right)$ with the help of an additional ancilla qubit, we obtain the HAM-T encoding of $\xi {C}$ such that 
\begin{equation}
    \bra{0}_a \text{HAM-T}_{\xi C,m} \ket{0}_a = \sum_{j=0}^{M-1}\sum_{l=0}^{M_H-1} \ket{j}\bra{j} \otimes \ket{l}\bra{l} \otimes \frac{\xi_j {C}(mh+lh/M_H)}{\Xi \alpha_C}. 
\end{equation}

To construct the block-encoding of $e^{-i\xi B t}$, we write 
\begin{equation}
\begin{split}
    e^{-i\xi_j B (mh+lh/M_H)} &= e^{-i(-\Xi+2j\Xi/M) B (mh+lh/M_H)} \\
    & = e^{i \Xi B mh} \left(e^{-i (2\Xi/M) B mh} \right)^j \left(e^{i\Xi B h/M_H}\right)^l \left( \left( e^{-i (2\Xi/M)B h/M_H}\right)^j \right)^l. 
\end{split}
\end{equation}
Noticing that the matrix $B$ can be diagonalized with QFT and the diagonal components are given through the oracle $O_1$, we can implement $e^{-i B s}$ for any real number $s$ fast-forwardly with $\mathcal{O}(1)$ uses of $O_1$, controlled phase gate and QFT. 
Then, according to the binary encoding of $0 \leq j \leq M-1$, we can use the controlled version of a total of $\log_2(M)$ operators $e^{-i (2\Xi/M) B mh}$, $e^{-i 2(2\Xi/M) B mh}$, $e^{-i 2^2(2\Xi/M) B mh}$, $\cdots$, $e^{-i 2^{\log_2(M)}(2\Xi/M) B mh}$ to implement the controlled evolution $\sum_{j=0}^{M-1} \ket{j}\bra{j} \otimes \left(e^{-i (2\Xi/M) B mh} \right)^j$. 
Similarly, we can construct the evolution $\sum_{l=0}^{M_H-1} \ket{l}\bra{l} \otimes \left( e^{i\Xi B h/M_H} \right)^l$ and $\sum_{j=0}^{M-1}\sum_{l=0}^{M_H-1} \ket{j}\bra{j} \otimes \ket{l}\bra{l} \otimes \left( \left( e^{-i (2\Xi/M)B h/M_H}\right)^j \right)^l$ with logarithmic cost as well. 
Multiplying them together gives the select oracle 
\begin{equation}
    \text{SEL}_{B,m} =  \sum_{j=0}^{M-1}\sum_{l=0}^{M_H-1} \ket{j}\bra{j} \otimes \ket{l}\bra{l} \otimes e^{-i\xi_j B (mh+lh/M_H)}. 
\end{equation}
Then 
\begin{equation}
    \text{HAM-T}_{H_I,m} := (I_{n_a} \otimes \text{SEL}_{B,m}^{\dagger}) \text{HAM-T}_{\xi C, m} (I_{n_a} \otimes \text{SEL}_{B,m}) 
\end{equation}
gives the HAM-T encoding of $H_I$ that 
\begin{equation}
    \bra{0}_a \text{HAM-T}_{H_I,m} \ket{0}_a = \sum_{j=0}^{M-1}\sum_{l=0}^{M_H-1} \ket{j}\bra{j} \otimes \ket{l}\bra{l} \otimes \frac{H_I(mh+lh/M_H;\xi_j)}{\Xi \alpha_C}. 
\end{equation}

The $\text{HAM-T}_{H_I,m}$ serves as the Hamiltonian input oracle in the truncated Dyson series method. 
Therefore, the method in~\cite{LowWiebe2019} gives the select oracle 
\begin{equation}
    \text{SEL}_W = \sum_{j=0}^{M-1} \ket{j}\bra{j} \otimes W_j, 
\end{equation}
where $W_j$ is an approximation of $U_I(T;\xi_j)$. 
We then multiply it on the left by $\sum_{j=0}^{M-1} \ket{j}\bra{j} \otimes e^{-i\xi_j BT}$ (which again can be efficiently constructed according to the binary representation of $j$) and obtain 
\begin{equation}
    \text{SEL}_{U} =  \sum_{j=0}^{M-1} \ket{j}\bra{j} \otimes U_j. 
\end{equation}
Here $U_j$ is an approximation of $e^{-i\xi_j B T} \mathcal{T} e^{-i \int_0^T H_I(s;\xi_j) ds}$. 
The $\text{SEL}_{U}$ operator serves as the select oracle in the LCU subroutine. 
Hence the formula~\cref{eqn:LCHS_discrete} can be directly implemented by the LCU technique (\cref{lem:LCU}). 

The overall complexity of the algorithm is given as follows. 

\begin{thm}\label{thm:LCHS-IP_complexity}
    Consider the spatially discretized equation~\cref{eqn:FRDE_linear_ODE_shifted}. 
    Suppose that 
    \begin{enumerate}
        \item we are given oracles encoding the eigenvalues of $B$ and the diagonal entries of ${C}(t)$, \emph{i.e.},  $O_1$ and $O_2$ defined in~\cref{eqn:def_O_high_d_linear} and~\cref{eqn:def_O_high_d_potential}, and the state preparation oracle $O_u: \ket{0} \rightarrow \ket{\vec{u}_0}$, 
        \item $u(t,x)$ is in the Gevrey class $G^{\sigma}$ in the sense that $u(t,x)$ is smooth and there exist constants $\Lambda > 0$ and $\sigma \geq 0$ such that 
    \begin{equation}
        \sup_{j\in[d],t\in[0,T],x\in[0,1]^d}|\partial_{x_j}^p u(t,x)| \leq \Lambda^{p+1} (p!)^{\sigma}, 
    \end{equation}
    \item $\|(u(T,n/N))_{n\in[N]^d}\| \geq \widetilde{g}(T)$ for a function $\widetilde{g}$,  
     \item $\|\vec{u}(0)\|/\|\vec{u}(T)\| \leq g(T)$ for a function $g$. 
    \end{enumerate}
    Then, the linear combination of Hamiltonian simulation in the interaction picture can prepare an $\epsilon$-approximation of $\ket{(u(T,n/N))_{n\in [N]^d}}$ using 
    \begin{enumerate}
        \item queries to $O_1$ and $O_2$ a total number of times 
        \begin{equation}
            \mathcal{O}\left(g(T)^2 \frac{T}{\epsilon} \log^3\left(\frac{g(T)T d\log d  }{\epsilon}   \log\left(\frac{1}{\widetilde{g}(T)}\right)   \right)\right), 
        \end{equation}
        \item queries to $O_u$ for $\mathcal{O}(g(T))$ times, 
        \item additional elementary gates for 
        \begin{equation}
            \mathcal{O}\left(g(T)^2 \frac{T}{\epsilon} \log\left(\frac{g(T) T}{\epsilon}\right) \left(d\log^2\left( d +  \log\left(\frac{T}{\widetilde{g}(T)}\right) + \log\left(\frac{1}{\epsilon}\right) \right) + \log\left( \frac{g(T) T }{\epsilon} \right) \right) \right).  
        \end{equation}
    \end{enumerate}
\end{thm}
\begin{proof}
    As discussed before, the oracle $\text{HAM-T}_{\xi C,m}$ can be implemented with $\mathcal{O}(1)$ queries to $O_2$ and $\mathcal{O}(d\log(N))$ additional gates~\cite[Lemma 48]{GilyenSuLowEtAl2019}, and the construction of $\text{SEL}_{B,m}$ requires $\mathcal{O}(\log(M)\log(M_H))$ $O_1$ and QFT (which requires $\mathcal{O}(d\log^2(N))$ gates). 
    So each $\text{HAM-T}_{H_I,m}$ can be constructed with $\mathcal{O}(\log(M)\log(M_H))$ queries to $O_1$, $O_2$ and $\mathcal{O}(d\log^2(N))$ additional gates. 
    According to~\cite[Corolllary 4]{LowWiebe2019}, by choosing 
    \begin{equation}
    \begin{split}
        M_H &= \mathcal{O}\left( \frac{T}{\Xi \alpha_C \epsilon'}  \left(  \|B\|\Xi^2\alpha_C + \Xi \max\|{C}'\| + \Xi^2 \alpha_C^2 \right)\right) \\
        & = \mathcal{O}\left( \frac{T}{ \epsilon'}  \left(  \|B\|\Xi +  \max\|{C}'\|\right)\right), 
    \end{split}
    \end{equation}
    we may implement the select oracle $\text{SEL}_W = \sum_{j=0}^{M-1} \bra{j}\ket{j}\otimes W_j$ such that 
    \begin{equation}
        \left\| W_j - \mathcal{T} e^{-i \int_0^T H_I(s;\xi_j) ds} \right\| \leq \epsilon'
    \end{equation}
    with 
    \begin{equation}
        \mathcal{O}\left(\Xi \alpha_C T \frac{\log(\Xi \alpha_C T/\epsilon')}{\log\log(\Xi \alpha_C T/\epsilon')}\right) = \mathcal{O}\left(\Xi T \frac{\log(\Xi T/\epsilon')}{\log\log(\Xi T/\epsilon')}\right)
    \end{equation}
    queries to $\text{HAM-T}_{H_I,m}$ and 
    \begin{equation}
    \begin{split}
        & \quad \mathcal{O}\left( \Xi \alpha_C T \frac{\log(\Xi \alpha_C T/\epsilon')}{\log\log(\Xi \alpha_C T/\epsilon')} \left(d\log(N) + \log(M_H) \right) \right) \\
        & = \mathcal{O}\left( \Xi T \frac{\log(\Xi T/\epsilon')}{\log\log(\Xi  T/\epsilon')} \left(d\log(N) + \log(M_H) \right) \right) 
    \end{split}
    \end{equation}
    additional gates. 
    Hence the $\text{SEL}_U$ can be implemented with asymptotically the same cost, and, taking into account the cost of constructing $\text{HAM-T}_{H_I,m}$, this step needs 
    \begin{equation}
        \mathcal{O}\left(\Xi T \frac{\log(\Xi T/\epsilon')}{\log\log(\Xi T/\epsilon')}\log(M)\log(M_H)\right) 
    \end{equation}
    queries to $O_1$ and $O_2$, and 
    \begin{equation}
        \mathcal{O}\left( \Xi T \frac{\log(\Xi T/\epsilon')}{\log\log(\Xi  T/\epsilon')} \left(d\log^2(N) + \log(M_H) \right) \right) 
    \end{equation}
    additional gates. 

    The LCU algorithm requires a single application of the select oracle and two applications of the prepare oracle for $\frac{1}{\sqrt{\|w\|_1}}\sum_{j=0}^{M-1} \sqrt{w_j}\ket{j}$. 
    Noticing that $w$ represents the discretized Cauchy distribution, we can implement this prepare oracle with $\mathcal{O}(\log(M))$ gates~\cite{GroverRudolph2002} and $\|w\|_1 = \mathcal{O}(1)$. 
    The output of the LCU step can be written as $\frac{1}{\|w\|\|\vec{u}_0\|} \ket{0} \widetilde{v} + \ket{\perp}$, where 
    \begin{equation}
        \widetilde{v} = \sum_{j=0}^{M-1} w_j U_j \vec{u}_0. 
    \end{equation}
    Using the inequality $\|\vec{a}/\|a\| - \vec{b}/\|b\|\| \leq 2\|\vec{a}-\vec{b}\|/\|\vec{a}\|$, we can bound the error in the quantum state as 
    \begin{equation}
        \begin{split}
            \|\ket{\vec{u}(T)} - \ket{\widetilde{v}}\| & \leq \frac{2}{\|\vec{u}(T)\|} \|\vec{u}(T) - \ket{\widetilde{v}}\| \\
            & \leq \frac{2 \|\vec{u}_0\|}{\|\vec{u}(T)\|} \left\|  \mathcal{T} e^{ - \int_0^T (B+{C}(s)) ds } - \sum_{j=0}^{M-1} w_j e^{-i\xi_j B T} \mathcal{T} e^{-i \int_0^T H_I(s;\xi_j) ds} \right\| \\
            & \quad\quad + \frac{2 \|\vec{u}_0\|}{\|\vec{u}(T)\|} \left\|  \sum_{j=0}^{M-1} w_j e^{-i\xi_j B T} \mathcal{T} e^{-i \int_0^T H_I(s;\xi_j) ds} - \sum_{j=0}^{M-1} w_j U_j \right\| \\
            & \leq \frac{2 \|\vec{u}_0\|}{\|\vec{u}(T)\|} \left\|  \mathcal{T} e^{ - \int_0^T (B+{C}(s)) ds } - \sum_{j=0}^{M-1} w_j e^{-i\xi_j B T} \mathcal{T} e^{-i \int_0^T H_I(s;\xi_j) ds} \right\| + \frac{2 \|\vec{u}_0\|}{\|\vec{u}(T)\|} \|w\|_1 \epsilon'. 
        \end{split}
    \end{equation}
    To bound the error by $\epsilon$, according to~\cref{lem:LCHS_discretization_error}, it suffices to choose 
    \begin{equation}
        \Xi = \mathcal{O}\left( \frac{\|\vec{u}_0\|}{\|\vec{u}(T)\|}\frac{1}{\epsilon} \right), \quad M = \mathcal{O}\left( \left(\frac{\|\vec{u}_0\|}{\|\vec{u}(T)\|}\right)^3\frac{T\|B\|}{\epsilon^3} \right), \quad \epsilon' = \mathcal{O} \left( \frac{\|\vec{u}(T)\|}{\|\vec{u}_0\|} \epsilon \right). 
    \end{equation}
    With these and by $\|B\| = \mathcal{O}(d^{\alpha/2}N^{\alpha})$ and $N$ given in~\cref{cor:FRDE_potential_spatial_N}, in each run of the algorithm we need queries to $O_u$ for $\mathcal{O}(1)$ times, queries to $O_1$ and $O_2$ for 
    \begin{equation}
    \begin{split}
        & \quad \mathcal{O}\left(\Xi T \frac{\log(\Xi T/\epsilon')}{\log\log(\Xi T/\epsilon')}\log(M)\log(M_H)\right) \\
        & = \mathcal{O}\left(g(T) \frac{T}{\epsilon} \log^3\left(\frac{g(T)T d\log d  }{\epsilon}   \log\left(\frac{1}{\widetilde{g}(T)}\right)   \right)\right),  
    \end{split}
    \end{equation}
    and additional gates for a total number of 
    \begin{equation}
    \begin{split}
        & \quad \mathcal{O}\left( \log(M) + \Xi T \frac{\log(\Xi T/\epsilon')}{\log\log(\Xi  T/\epsilon')} \left(d\log^2(N) + \log(M_H) \right) \right) \\
        & = \mathcal{O}\left(g(T) \frac{T}{\epsilon} \log\left(\frac{g(T) T}{\epsilon}\right) \left(d\log^2\left( d +  \log\left(\frac{T}{\widetilde{g}(T)}\right) + \log\left(\frac{1}{\epsilon}\right) \right) + \log\left( \frac{g(T) T }{\epsilon} \right) \right) \right). 
    \end{split}
    \end{equation}
    With amplitude amplification, the average number of repeats to get a success is $\mathcal{O}(\|w\|\|\vec{u}_0\|/\|\widetilde{v}\|) = \mathcal{O}(g(T))$, so the overall complexity should be multiplied by this factor. 
\end{proof}

\section{Nonlinear equations}\label{sec:nonlinear}

We now discuss the full non-linear fractional reaction-diffusion equation as in~\cref{eqn:FRDE}. 
For simplicity we only consider time-independent potential function $c(t,x) \equiv c(t)$. 
After spatial discretization,~\cref{eqn:FRDE} becomes a system of nonlinear ODEs with quadratic nonlinear term, which can be tackled by the Carleman linearization technique~\cite{LiuKoldenKroviEtAl2021}. 
Existing algorithms based on Carleman linearization assume the sparsity of the coefficient matrices in the nonlinear ODEs. 
This facilities the construction of the Carleman matrix, which is a direct sum of matrices in different dimensions. 
However, in the fractional reaction-diffusion equation, the coefficient matrices are unavoidably dense, so we will discuss a block-encoding implementation of the Carleman linearization technique. 

\subsection{Spatial discretization}

We use the same spatial discretization as in the linear case. 
Let $\vec{u}(t)$ be a $N^d$-dimensional vector approximating the exact solution $u(t,x)$ at equi-distant grid points $(n_1/N.\cdots,n_d/N)$ where $n_j \in [N]$. 
The spatially discretized equation can be written as 
\begin{equation}\label{eqn:FRDE_ODE_nonlinear}
    \frac{d}{dt} \vec{u} = F_1 \vec{u} + F_2 \vec{u}^{\otimes 2}. 
\end{equation}
Here $\vec{u}^{\otimes 2} = \vec{u} \otimes \vec{u}$ is a $N^{2d}$-dimensional vector. 
$F_1 = -B - C + aI$. 
$F_2$ is a $N^{d} \times N^{2d}$ dimensional matrix that maps $\vec{u}^{\otimes 2}$ to $-a \vec{u}$, \emph{i.e.}, each row of $F_2$ only has one non-zero entry to be $-a$ at its $((j-1)N^d+j)$-th column (for the $j$-th row).

\subsection{Carleman linearization}

The idea of Carleman linearization for~\cref{eqn:FRDE_ODE_nonlinear} is to convert the nonlinear ODE to an equivalent infinite-dimensional linear ODE. 
Specifically, for any positive integer $m$, the tensor product $\vec{u}^{\otimes m}$ satisfies the ODE 
\begin{equation}
    \frac{d}{dt} (\vec{u}^{\otimes m}) = A_m^m \vec{u}^{\otimes m} + A_{m+1}^m \vec{u}^{\otimes (m+1)},
\end{equation}
where ($I$ represents the identity matrix of dimension $N^d \times N^d$)
\begin{equation}
    A_m^m = \sum_{j=1}^m I^{\otimes (j-1)} \otimes F_1 \otimes I^{\otimes (m-j)}, 
\end{equation}
\begin{equation}
    A_{m+1}^m = \sum_{j=1}^m I^{\otimes (j-1)} \otimes F_2 \otimes I^{\otimes (m-j)}. 
\end{equation}
So the infinite-dimensional vector $[\vec{u};\vec{u}^{\otimes 2};\vec{u}^{\otimes 3};\cdots]$ satisfies a system of homogeneous linear ODE with coefficient matrix 
\begin{equation}
    \left(  \begin{array}{cccccc}
        A_1^1 & A_2^1 & & & & \\
         & A_2^2 & A_3^2 & & & \\
         & & \ddots & \ddots & & \\
         & & & A_M^M & A_{M+1}^M & \\
         & & & & \ddots &\ddots \\
    \end{array}   \right). 
\end{equation}
To implement the Carleman linearization numerically, we truncate the infinite-dimensional ODE at a specific order $M$ and consider the ODE  
\begin{equation}\label{eqn:FRDE_Carleman_ODE}
    \frac{d}{dt} \vec{w} = A \vec{w}. 
\end{equation}
Here $\vec{w} = [\vec{w}_1;\vec{w}_2;\cdots;\vec{w}_M]$ and each $\vec{w}_j$ is an $N^{jd}$-dimensional vector expected to approximate $\vec{u}^{\otimes j}$. 
The matrix $A$ can be represented as 
\begin{equation}
    A = \left(  \begin{array}{ccccc}
        A_1^1 & A_2^1 & & & \\
         & A_2^2 & A_3^2 & & \\
         & & \ddots & \ddots & \\
         & & & A_{M-1}^{M-1} & A_{M}^{M-1} \\
         & & & & A_M^M \\
    \end{array}   \right). 
\end{equation}

A crucial theoretical aspect on the efficiency of the Carleman linearization method is its convergence, i.e., whether or not $\vec{w}_1$, the first block in the solution of the truncated system in~\cref{eqn:FRDE_Carleman_ODE}, is a good approximation of $\vec{u}$, the solution of the nonlinear ODE in~\cref{eqn:FRDE_ODE_nonlinear}. Several existing works~\cite{ForetsPouly2017,AminiSunMotee2021,AminiZhengSunMotee2025,LiuKoldenKroviEtAl2021,AnFangJordanEtAl2022,CostaPhilippMoralesBerry2025,WuWangLi2025} have studied this topic and established rigorous convergence guarantee in different scenarios. 
For example, in the ODE case,~\cite{LiuKoldenKroviEtAl2021} shows that, when all the eigenvalues of the linear part $F_1$ are negative and the nonlinear part $F_2$ is relatively small compared to the decay rate of the linear part, $\vec{w}_1$ can be a good approximation of $\vec{u}$, and the truncation order $M$ only needs to be logarithmic in the precision. 

However, as pointed out in~\cite{AnFangJordanEtAl2022,CostaPhilippMoralesBerry2025}, such a convergence condition might be too strong or even unrealistic when $N$ is large. 
Then, for standard reaction-diffusion equations (i.e., $\alpha = 2$), these two works show an improved convergence analysis of the Carleman linearization with broader condition, by leveraging a maximum norm analysis~\cite{AnFangJordanEtAl2022,CostaPhilippMoralesBerry2025} and a rescaling technique~\cite{CostaPhilippMoralesBerry2025}. 
We expect the same convergence for the fractional order as well. 
This is because the only difference between standard and fractional reaction diffusion equations lies in the eigenvalues of the discretized Laplace $B$, stretching from $\lambda$ to $\lambda^{\alpha/2}$. 
Therefore, we expect that one can prove the convergence of the Carleman linearization following the analysis in~\cite{AnFangJordanEtAl2022,CostaPhilippMoralesBerry2025} with a modified definition of the eigenvalues of $B$, and we leave a careful study with explicit technical statements for future work.

\subsection{Block-encoded implementation}

The coefficient matrix $A$ is expressed as a partitioned matrix with blocks in different dimension. 
When all the blocks are sparse, as assumed in existing quantum algorithm based on Carleman linearization, the entire matrix $A$ is still sparse and one can directly implement it from sparse input model. 
This is also true for (integer-order) reaction-diffusion equation when $\alpha = 2$. 
However, when $0 < \alpha < 2$, while the off-diagonal blocks $A_{m+1}^m$ are still sparse, the diagonal blocks $A_{m}^m$ are unavoidably dense. 
Though the block-encoding of each block $A_m^m$ is still construable, it is somewhat cumbersome to assemble them together to the block-encoding of the entire $A$. 

Inspired by the technique of~\cite{LiuKoldenKroviEtAl2021} for state preparation, we further enlarge the dimension of the linearized ODE. 
The purpose is to make the resulting enlarged coefficient matrix expressed as a partitioned matrix with blocks in the same dimension, and meanwhile a subspace of the enlarged solution is still the solution of the original ODE. 
Specifically, let us consider the ODE 
\begin{equation}\label{eqn:FRDE_Carleman_ODE_enlarged}
    \frac{d}{dt} \vec{y} = \widetilde{A} \vec{y}. 
\end{equation}
Here $\vec{y} = [\vec{y}_1;\cdots; \vec{y}_M]$ and each $\vec{y}_j$ is an $N^{Md}$-dimensional vector. 
The initial value $\vec{y}(0)$ is chosen to be $[\ket{0}^{\otimes M-1}\otimes \ket{\vec{u}(0)}; \ket{0}^{\otimes M-2}\otimes \ket{\vec{u}(0)}^{\otimes 2}; \cdots; \ket{\vec{u}(0)}^{\otimes M}]$ where $\ket{0} = (0,\cdots,0,1)^{T}$ in $N^d$ dimension. 
The coefficient matrix is 
\begin{equation}
    \widetilde{A} = \left(  \begin{array}{ccccc}
        \widetilde{A}_1^1 & \widetilde{A}_2^1 & & & \\
         & \widetilde{A}_2^2 & \widetilde{A}_3^2 & & \\
         & & \ddots & \ddots & \\
         & & & \widetilde{A}_{M-1}^{M-1} & \widetilde{A}_{M}^{M-1} \\
         & & & & \widetilde{A}_M^M \\
    \end{array}   \right). 
\end{equation}
Here each $\widetilde{A}_{m}^m$ and $\widetilde{A}_{m+1}^m$ is an $N^{Md}$-dimensional square matrix.  
$\widetilde{A}_m^m = I^{\otimes (M-m)} \otimes A_m^m$ for $m < M$ and $\widetilde{A}_M^M = A_M^M$. 
$\widetilde{A}_{m+1}^m$ contains $A_{m+1}^m$ at its most bottom right and otherwise $0$. 

We now show that the non-zero entries of $\vec{y}$ exactly form $\vec{w}$. 
To this end, for each $\vec{y}_m$, we write it as $[\vec{y}_{m,1};\cdots;\vec{y}_{m,N^{(M-m)d}}]$, where each $\vec{y}_{m,j}$ is an $N^{md}$-dimensional vector. 
By definition of $\vec{y}(0)$, we have that for every $m$, all but the last component $\vec{y}_{m,N^{(M-m)d}}(0)$ of $\vec{y}_m(0)$ are zero. 
Furthermore, by the definition of $\widetilde{A}$, the variables $\vec{y}_{m,j}$ for $j \neq N^{(M-m)d}$ does not interact with other variables outside. 
Therefore, $\vec{y}_{m,j}(t)$ is always $0$ for all $t$, $m$ and $j \neq N^{(M-m)d}$. 
Now, if we only focus on the ODEs that $\vec{y}_{m,N^{(M-m)d}}$'s satisfy, it is exactly the original ODE~\cref{eqn:FRDE_Carleman_ODE} with the same initial condition. 
So we have $[\vec{y}_{1,N^{(M-1)d}}(t);\vec{y}_{2,N^{(M-2)d}}(t);\cdots;\vec{y}_{M,1}(t)] = \vec{w}(t)$ for all $t$.  
This implies that, instead of solving~\cref{eqn:FRDE_Carleman_ODE}, we can focus on its equivalent formalism~\cref{eqn:FRDE_Carleman_ODE_enlarged}. 

\cref{eqn:FRDE_Carleman_ODE_enlarged} can be solved by standard quantum algorithm for linear ODEs. 
For example, we can use the method based on truncated Dyson series~\cite{BerryCosta2022} discussed in the previous section. 
To use this method, we need the state preparation for $\vec{y}(0)$ and the block-encoding of $\widetilde{A}$. 
The state preparation oracle can be constructed in a similar manner as in~\cite{LiuKoldenKroviEtAl2021}. 
To construct the block-encoding of $\widetilde{A}$, we decompose it as $\widetilde{A} = \widetilde{D} + \widetilde{R}$, where 
\begin{equation}
    \widetilde{D} = \left(  \begin{array}{ccccc}
        \widetilde{A}_1^1 &  & & & \\
         & \widetilde{A}_2^2 &  & & \\
         & & \ddots &  & \\
         & & & \widetilde{A}_{M-1}^{M-1} &  \\
         & & & & \widetilde{A}_M^M \\
    \end{array}   \right), \quad \widetilde{R} = \left(  \begin{array}{ccccc}
        0 & \widetilde{A}_2^1 & & & \\
         & 0 & \widetilde{A}_3^2 & & \\
         & & \ddots & \ddots & \\
         & & & 0 & \widetilde{A}_{M}^{M-1} \\
         & & & & 0\\
    \end{array}   \right). 
\end{equation}
Notice that $\widetilde{R}$ is a $\mathcal{O}(M)$-sparse matrix. 
According to~\cite[Lemma 48]{GilyenSuLowEtAl2019}, we may implement a block-encoding of $\mathcal{R}$ with $\mathcal{O}(1)$ query complexity and $\mathcal{O}(M)$ block-encoding factor. 
For $\widetilde{D}$, we start with the block-encoding of $F_1$, which can be constructed from the linear combination of $B,C$ and $aI$. 
According to~\cref{lem:LCU}, the block-encoding factor is $\mathcal{O}(d^{\alpha}N^{2\alpha})$, corresponding to the spectral norm of $B$, and the query complexity for block encoding $F_1$ is $\mathcal{O}(1)$. 
Denote this block-encoding by $U_{F_1}$. 
Then a block-encoding of $I^{\otimes (j-1)} \otimes F_1 \otimes I^{\otimes (m-j)}$ can be constructed by applying $U_{F_1}$ on the correct register. 
Since $A_m^m$ is the summation of $I^{\otimes (j-1)} \otimes F_1 \otimes I^{\otimes (m-j)}$, according to~\cref{lem:LCU}, we may further construct the block-encoding of $A_m^m$, denoted by $U_{A_m^m}$, with $\mathcal{O}(Md^{\alpha}N^{2\alpha})$ block-encoding factor. 
Here we choose the block-encoding factor for all $A_m^m$ to be the same and corresponds to the worst case $A_M^M$ with most summation terms in order to facilitate later construction for bigger matrix. 
Then, for each $m$, a block-encoding of $\widetilde{A}_m^m$, denoted by $U_{\widetilde{A}_m^m}$, can be constructed with a single use of $U_{A_m^m}$, and the block-encoding of $\widetilde{D}$ is given by 
\begin{equation}
    \sum_{m=0}^{M-1} \ket{m}\bra{m} \otimes U_{\widetilde{A}_m^m}. 
\end{equation}
This can be implemented by controlled version of $U_{A_m^m}$ and requires $\mathcal{O}(M)$ query complexity, and the block-encoding factor is $\mathcal{O}(Md^{\alpha}N^{2\alpha})$. 
The final step is to use~\cref{lem:LCU} again and, from the block-encoding of $\widetilde{D}$ and $\widetilde{R}$, we may construct the desired block-encoding of $\widetilde{A}$ with $\mathcal{O}(M)$ query complexity and $\mathcal{O}(Md^{\alpha}N^{2\alpha})$ block-encoding factor. 
Notice that both query complexity and the block-encoding factor does not involve exponential dependence on $d$. 

With the state preparation for $\vec{y}(0)$ and the block-encoding of $\widetilde{A}$, we can solve~\cref{eqn:FRDE_Carleman_ODE_enlarged} efficiently using truncated Dyson series method~\cite{BerryCosta2022}. 
Notice that the method requires $\widetilde{A}$ to have non-positive logarithmic norm, which can be guaranteed if all the eigenvalues of $F_1$ are negative and $F_2$ is bounded, which corresponds to the standard assumption on the boundedness of the nonlinearity (namely the condition $R_D < 1$ in~\cite{AnFangJordanEtAl2022}).

\section{Conclusion}\label{sec:conclusion}

In this paper, we study efficient quantum algorithms for linear and nonlinear fractional reaction-diffusion equations. 
For linear equations, we improve and analyze the complexity of four different methods: second-order Trotter formula, time-marching method, truncated Dyson series method, and the linear combination of Hamiltonian simulation with the interaction picture formalism (LCHS-IP). 
Among all the methods, the LCHS-IP method achieves best scaling in the spatial dimension and thus is most suiable for high-dimensional linear fractional reaction-diffusion equations. 
For nonlinear equations, we generalize the quantum Carleman linearization algorithm to the case with block-encoding input oracle, making the algorithm applicable to dense coefficient matrices. 

A natural extension of this work is to design better quantum algorithm for linear reaction-diffusion equations with near-optimal scalings in all parameters. 
A desired algorithm is expected to simultaneously scale poly-logarithmically in dimension (as LCHS-IP), poly-logarithmically in precision (as time-marching and truncated Dyson series methods), linearly in evolution time (as truncated Dyson series and LCHS-IP), and have low state preparation cost (as Trotter, time-marching and LCHS-IP). 
Among all the methods being considered in this paper, the LCHS-IP method is the closest one as it only misses the poly-logarithmic dependence on precision, so it is interesting to explore whether the LCHS-IP method can be further exponentially improved in terms of precision. 
Another possibility is to use higher-order product formula, whose asymptotic scaling tends to be $
T^{1+o(1)}/\epsilon^{o(1)}$. 
However, as proved in~\cite{ChildsSuTranEtAl2020}, there might be extra exponential overhead in Trotter errors when we deal with non-unitary dynamics. 
Therefore a tailored design of the product formula and an improved error analysis would be necessary. 
Furthermore, to obtain a poly-logarithmic dependence on dimension, we may need to take advantage of the vector-norm scaling of the product formula, which states that the Trotter errors may be independent of the spectral norm of the Hamiltonians if the quantum states are within a more regular subspace with better smoothness assumption. 
Such a vector-norm scaling has been proved for first- and second-order Trotter applied to Hamiltonian simulation in~\cite{AnFangLin2021}, and it remains open to establish similar error bounds for higher-order product formula and the cases beyond Hamiltonian simulation. 
This is our ongoing work. 

For non-linear equations, a natural next step is to establish a rigorous analysis with detailed computational costs. 
We expect the analysis presented in~\cite{AnFangJordanEtAl2022} to work with suitable modifications. 
However, the complexity estimate in~\cite{AnFangJordanEtAl2022} still depends polynomially on the dimension $d$. 
It is interesting to explore whether the quantum Carleman algorithm can also avoid such a polynomial overhead with the help of tighter error bounds, or new techniques are necessary to achieve this task. 

Throughout this paper, we focus on the spectral fractional Laplacian with periodic boundary condition, which facilitates its spatial discretization and quantum implementation of its time evolution through the QFT circuit. 
Our future work will be focusing on the fractional Laplacian operator with Riesz definition and exploring the efficiency of quantum algorithms.

\section*{Acknowledgments} 

DA acknowledges the support by the Department of Defense through the Hartree Postdoctoral Fellowship at QuICS, and the seed grant at the NSF Quantum Leap Challenge Institute for Robust Quantum Simulation (QLCI grant OMA-2120757). KT gratefully acknowledges the support by the National Science Foundation under the grants DMS-2231533 and DMS-2008568.

\bibliographystyle{quantum}
\bibliography{main}

\clearpage
\appendix

\section{Technical lemmas}

\begin{lem}\label{lem:error_vector_quantum_state}
    Let $\vec{a}$ and $\vec{b}$ be two non-zero vectors, possibly unnormalized.
    Then 
    \begin{equation}
        \left\|\frac{\vec{a}}{\|\vec{a}\|}-\frac{\vec{b}}{\|\vec{b}\|}\right\| \leq \frac{2\|\vec{a}-\vec{b}\|}{\|\vec{a}\|}. 
    \end{equation}
\end{lem}

\begin{lem}\label{lem:variation_of_parameters}
    Suppose $A(t)$ is a matrix-valued continuous function and the operator $S(t,s)$ solves the differential equation 
    \begin{equation}
        \frac{d}{dt} S(t,s) = A(t) S(t,s), \quad S(s,s) = I. 
    \end{equation}
    Then, 
    \begin{enumerate}
        \item for any matrix-valued continuous function $R(t)$, the solution of the differential equation 
    \begin{equation}
        \frac{d}{dt} \widetilde{S}(t,0) = A(t) \widetilde{S}(t,0) + R(t), \quad \widetilde{S}(0,0) = I 
    \end{equation}
    can be represented as 
    \begin{equation}
        \widetilde{S}(t,0) = S(t,0) + \int_0^t S(t,s) R(s) ds. 
    \end{equation}
        \item for any vector-valued continuous function $\vec{r}(t)$, the solution of the differential equation 
    \begin{equation}
        \frac{d}{dt} \vec{\psi}(t) = A(t) \vec{\psi}(t) + \vec{r}(t), \quad \vec{\psi}(0) = \vec{\psi}_0
    \end{equation}
    can be represented as 
    \begin{equation}
        \vec{\psi}(t) = S(t,0)\vec{\psi}_0 + \int_0^t S(t,s) \vec{r}(s) ds. 
    \end{equation}
    \end{enumerate}
\end{lem}

\section{Proof of\texorpdfstring{~\cref{thm:fractional_heat}}{}}\label{app:proof_linear_without_potential}

\begin{proof}
    For any function $f$ defined on $[0,1]^d$, its Fourier coefficient is defined to be
    \begin{equation}
        \hat{f}_k = \int_{[0,1]^d} f(x) e^{-2\pi i (k_0x_0+\cdots+k_{d-1}x_{d-1})} dx. 
    \end{equation}
    Using integration by parts for $p$ times, we may obtain 
    \begin{equation}\label{eqn:Fourier_coefficient_bound}
        |\hat{f}_k| \leq \frac{\max_{j}\|\partial_{x_j}^p f(x) \|_{L^1}}{(2\pi \|k\|_{\infty})^p}. 
    \end{equation}
    As a result, for each fixed $m$, 
    \begin{equation}
        \begin{split}
            \left|\sum_{j\in\mathbb{Z}^d} \hat{u}_{m+jN} - \hat{u}_{i(m)} \right| &\leq \sum_{l=1}^{\infty} \sum_{\|j\|_{\infty} = l} |\hat{u}_{i(m)+jN}| \\
            & \leq \sum_{l=1}^{\infty} ((l+1)^d-l^d) \frac{\max_{j}\|\partial_{x_j}^p u_0 (x) \|_{L^1}}{(\pi (2l-1)N)^p}. 
        \end{split}
    \end{equation}
    Suppose that $p \geq d+2$. 
    Using the inequality that $2l-1\geq l+1$ for all $l \geq 2$, we have 
    \begin{equation}
        \begin{split}
            \left|\sum_{j\in\mathbb{Z}^d} \hat{u}_{m+jN} - \hat{u}_{i(m)} \right| &\leq \left(2^d-1  + \sum_{l=2}^{\infty}  \frac{(l+1)^d-l^d}{(2l-1)^p}\right)\frac{\max_{j}\|\partial_{x_j}^p u_0(x) \|_{L^1}}{\pi^p N^p} \\
            & \leq \left(2^d-1  + \sum_{l=2}^{\infty}  \frac{1}{(2l-1)^{p-d}}\right)\frac{\max_{j}\|\partial_{x_j}^p u_0(x) \|_{L^1}}{\pi^p N^p} \\
            & \leq \left(2^d-1  + \int_1^{\infty} \frac{dx}{(2x-1)^{p-d}}\right)\frac{\max_{j}\|\partial_{x_j}^p u_0(x) \|_{L^1}}{\pi^p N^p} \\
            & =  \left(2^d-1  + \frac{1}{2(p-d-1)}\right)\frac{\max_{j}\|\partial_{x_j}^p u_0(x) \|_{L^1}}{\pi^p N^p} \\
            & \leq \frac{\max_{j}\|\partial_{x_j}^p u_0(x) \|_{L^1}}{(\pi/2)^p N^p}. 
        \end{split}
    \end{equation}
    Therefore, 
    \begin{equation}\label{eqn:Fourier_error_bound}
    \begin{split}
         & \quad \left\|(\mathcal{F}^{-1})^{\otimes d} \vec{u}_0 - N^{d/2} \sum_{m\in [N]^d} \hat{u}_{i(m)}  \ket{m_0}\cdots\ket{m_{d-1}}\right\| \\
         & = N^{d/2} \left\|\sum_{m\in [N]^d} \left(\sum_{j\in \mathbb{Z}^d} \hat{u}_{m+jN} -\hat{u}_{i(m)}\right) \ket{m_0}\cdots\ket{m_{d-1}}\right\| \\
         & \leq  \frac{\max_{j}\|\partial_{x_j}^p u_0(x) \|_{L^1}}{(\pi/2)^p N^{p-d}}.  
    \end{split}
    \end{equation}
    Similarly, 
    \begin{equation}
        \left\|(\mathcal{F}^{-1})^{\otimes d} \vec{u}(T,x) - N^{d/2} \sum_{m\in [N]^d} \hat{u}_{i(m)} e^{-(2\pi\|i(m)\|)^{\alpha} T} \ket{m_0}\cdots\ket{m_{d-1}}\right\| \leq \frac{\max_{j}\|\partial_{x_j}^p u(T,x) \|_{L^1}}{(\pi/2)^p N^{p-d}}. 
    \end{equation}
    Let $\widetilde{\mathcal{F}} = (\mathcal{F})^{\otimes d}\otimes I \otimes I$ and $U$ denote the composition of $O_1$, $O_{\exp,1}$ and c-$R$ specified in~\cref{fig:circuit_FRDE_linear}, then 
    \begin{equation}
    \begin{split}
        & \quad \left\| (I^{\otimes d}\otimes \bra{0}^{\otimes 3} )\widetilde{\mathcal{F}}U\widetilde{\mathcal{F}}^{-1} \ket{u_0}\ket{0}^{\otimes 3} - \frac{1}{\|\vec{u}_0\|} \vec{u}(T) \right\| \\
        & \leq \left\|  (I^{\otimes d}\otimes \bra{0}^{\otimes 3})\widetilde{\mathcal{F}}U\widetilde{\mathcal{F}}^{-1} \ket{u_0}\ket{0}^{\otimes 3} - (I^{\otimes d}\otimes \bra{0}^{\otimes 3})\widetilde{\mathcal{F}}U \frac{N^{d/2}}{\|\vec{u}_0\|} \sum_{m\in [N]^d} \hat{u}_{i(m)}  \ket{m_0}\cdots\ket{m_{d-1}}\ket{0}^{\otimes 3} \right\| \\
        & \quad + \left\| (I^{\otimes d}\otimes \bra{0}^{\otimes 3})\widetilde{\mathcal{F}}\left(\frac{N^{d/2}}{\|\vec{u}_0\|} \sum_{m\in [N]^d} \hat{u}_{i(m)}  e^{-(2\pi\|i(m)\|)^{\alpha}T} \ket{m_0}\cdots\ket{m_{d-1}}\ket{0}^{\otimes 3} + \ket{\perp} \right) \right. \\
        & \quad \quad \left. - \frac{1}{\|\vec{u}_0\|} \sum_{m \in [N]^d } u(T,m/N) \ket{m_0}\cdots\ket{m_{d-1}} \right\|  \\
        & \leq \left\| (\mathcal{F}^{-1})^{\otimes d} \ket{u_0} - \frac{N^{d/2}}{\|\vec{u}_0\|} \sum_{m\in [N]^d} \hat{u}_{i(m)}  \ket{m_0}\cdots\ket{m_{d-1}} \right\| \\
        & \quad + \left\| \mathcal{F}^{\otimes d} \frac{N^{d/2}}{\|\vec{u}_0\|} \sum_{m\in [N]^d} \hat{u}_{i(m)}  e^{-(2\pi\|i(m)\|)^{\alpha}T} \ket{m_0}\cdots\ket{m_{d-1}} -  \frac{1}{\|\vec{u}_0\|} \sum_{m \in [N]^d } u(T,m/N) \ket{m_0}\cdots\ket{m_{d-1}} \right\| \\
        & \leq \frac{1}{\|\vec{u}_0\|} \frac{\max_{j}\|\partial_{x_j}^p u_0(x) \|_{L^1} + \max_{j}\|\partial_{x_j}^p u(T,x) \|_{L^1} }{(\pi/2)^p N^{p-d}}. 
    \end{split}
    \end{equation}
    By~\cref{lem:error_vector_quantum_state}, the overall $2$-norm error of the output quantum state after successful measurement can be bounded by 
    \begin{equation}
        \frac{4}{\|\vec{u}_0\|} \frac{\max_{j,t\in\left\{0,T\right\}}\|\partial_{x_j}^p u(t,x) \|_{L^1}}{(\pi/2)^p N^{p-d}} \frac{\|\vec{u}_0\|}{\|\vec{u}(T)\|} = \frac{4 \max_{j,t\in\left\{0,T\right\}}\|\partial_{x_j}^p u(t,x) \|_{L^1}}{\|\vec{u}(T)\|(\pi/2)^p N^{p-d} }. 
    \end{equation}
    
    The claimed complexity can be shown by noticing that in each run of~\cref{fig:circuit_FRDE_linear}, we need $\mathcal{O}(1)$ queries to the aforementioned oracles, and $\mathcal{O}(d\log^2(N))$ gates due to QFT and controlled rotations, and under amplitude amplification, the averaged repeats to get a success is $\mathcal{O}(\|\vec{u}_0\|/\|\vec{u}(T)\|)$. 
\end{proof}

\section{Bounding the spatial discretization errors}\label{app:spatial_discretization_error}

Here we present the detailed proofs of the spatial discretization results (\cref{lem:FRDE_potential_spatial_error},~\cref{cor:FRDE_potential_spatial_error_exp} and~\cref{cor:FRDE_potential_spatial_N}). 

\begin{proof}[Proof of~\cref{lem:FRDE_potential_spatial_error}]
    Let $\vec{u}_e(t) = (u(t,n/N))_{n\in[N]^d}$ denote the exact solution evaluated at discrete grid points. 
    Using~\cref{eqn:FRDE_linear}, for any $n \in [N]^d$, 
    \begin{equation}
    \begin{split}
        \partial_t u(t,n/N) &= -(-\Delta)^{\alpha/2} u(t,n/N) - c(t,n/N) u(t,n/N) \\
        & = \left(-B\vec{u}_e\right)_n - c(t,n/N) u(t,n/N) + \left(B\vec{u}_e\right)_n - (-\Delta)^{\alpha/2} u(t,n/N). 
    \end{split}
    \end{equation}
    Therefore the vector $\vec{u}_e(t)$ solves the differential equation 
    \begin{equation}
        \frac{d}{dt} \vec{u}_e = -B\vec{u}_e - C(t) \vec{u}_e + \vec{r}(t), 
    \end{equation}
    where 
    \begin{equation}
        \left(\vec{r}(t)\right)_n = \left(B\vec{u}_e\right)_n - (-\Delta)^{\alpha/2} u(t,n/N). 
    \end{equation}
    Therefore, the equations $\vec{u}_e$ satisfies can be viewed as a perturbation (by $\vec{r}$) of the equations $\vec{u}$ satisfies. 
    By~\cref{lem:variation_of_parameters}, we have 
    \begin{equation}
        \vec{u}_e(t) = \vec{u}(t) + \int_0^t \mathcal{T} e^{\int_s^t (-B-C(\tau)) d\tau}  \vec{r}(s) ds, 
    \end{equation}
    and thus 
    \begin{equation}\label{eqn:discretization_error_intermediate}
        \|\vec{u}_e(T) - \vec{u}(T)\| \leq T \max_{t\in[0,T]} \|\vec{r}(t)\|. 
    \end{equation}
    
    It remains to bound $\|\vec{r}(t)\|$, which can be done similarly to the proof of~\cref{thm:fractional_heat}. 
    Suppose that the Fourier series of $u(t,x)$ is 
    \begin{equation}
        u(t,x) = \sum_{k\in \mathbb{Z}^d} \hat{u}_k(t) e^{2\pi i (k_0x_0+\cdots+k_{d-1}x_{d-1})}, 
    \end{equation}
    where 
    \begin{equation}
        \hat{u}_k(t) = \int_{[0,1]^d} u(t,x) e^{-2\pi i (k_0x_0+\cdots+k_{d-1}x_{d-1})} dx. 
    \end{equation}
    By the same reasoning of~\cref{eqn:Fourier_error_bound}, 
    \begin{equation}
        \left\|(\mathcal{F}^{-1})^{\otimes d} \vec{u}_e(t) - N^{d/2} \sum_{m\in [N]^d} \hat{u}_{i(m)}(t)  \ket{m_0}\cdots\ket{m_{d-1}}\right\|
         \leq  \frac{\max_{j}\|\partial_{x_j}^p u(t,x) \|_{L^1}}{(\pi/2)^p N^{p-d}}. 
    \end{equation}
    Noticing that $\|D\| \leq \pi^{\alpha} N^{\alpha} d^{\alpha/2}$, we have 
    \begin{equation}\label{eqn:eq1}
        \left\|D(\mathcal{F}^{-1})^{\otimes d} \vec{u}_e(t) - N^{d/2} \sum_{m\in [N]^d} 2^{\alpha} \pi^{\alpha} \|i(m)\|^{\alpha} \hat{u}_{i(m)}(t)  \ket{m_0}\cdots\ket{m_{d-1}}\right\|
         \leq  \frac{2^p d^{\alpha/2} \max_{j}\|\partial_{x_j}^p u(t,x) \|_{L^1}}{\pi^{p-\alpha} N^{p-d-\alpha}}.  
    \end{equation}
    Again using the derivation of~\cref{eqn:Fourier_error_bound} and noting that \\
    $(-\Delta)^{\alpha/2} u(t,x) = \sum_{k\in \mathbb{Z}^d} 2^{\alpha}\pi^{\alpha}\|k\|^{\alpha} \hat{u}_k(t) e^{2\pi i (k_0x_0+\cdots+k_{d-1}x_{d-1})}$, we have 
    \begin{equation}
    \begin{split}
        & \quad \left\|(\mathcal{F}^{-1})^{\otimes d} \left((-\Delta)^{\alpha/2} u(t,n/N)\right)_{n\in[N]^d} - N^{d/2} \sum_{m\in [N]^d} 2^{\alpha}\pi^{\alpha}\|i(m)\|^{\alpha} \hat{u}_{i(m)}(t)  \ket{m_0}\cdots\ket{m_{d-1}}\right\| \\
         &= N^{d/2} \left\|\sum_{m\in [N]^d} \left(\sum_{j\in \mathbb{Z}^d} 2^{\alpha}\pi^{\alpha}\|m+jN\|^{\alpha} \hat{u}_{m+jN}(t) -2^{\alpha}\pi^{\alpha}\|i(m)\|^{\alpha} \hat{u}_{i(m)}(t)\right) \ket{m_0}\cdots\ket{m_{d-1}}\right\|. 
    \end{split}
    \end{equation}
    Using~\cref{eqn:Fourier_coefficient_bound}, for each fixed $m$, we have 
    \begin{equation}
        \begin{split}
            & \quad \left|\sum_{j\in \mathbb{Z}^d} 2^{\alpha}\pi^{\alpha}\|m+jN\|^{\alpha} \hat{u}_{m+jN}(t) -2^{\alpha}\pi^{\alpha}\|i(m)\|^{\alpha} \hat{u}_{i(m)}(t)\right| \\
            & \leq \sum_{l=1}^{\infty} \sum_{\|j\|_{\infty} = l} 2^{\alpha}\pi^{\alpha}\|i(m)+jN\|^{\alpha} |\hat{u}_{i(m)+jN}(t)| \\
            & \leq \sum_{l=1}^{\infty} \sum_{\|j\|_{\infty} = l} 2^{\alpha}\pi^{\alpha}\|i(m)+jN\|^{\alpha} \frac{\max_{j}\|\partial_{x_j}^p u(t,x) \|_{L^1}}{(2\pi \|i(m)+jN\|_{\infty})^p} \\
            & \leq \sum_{l=1}^{\infty} ((l+1)^d-l^d) \pi^{\alpha} (2l+1)^{\alpha}N^{\alpha}d^{\alpha/2} \frac{\max_{j}\|\partial_{x_j}^p u(t,x) \|_{L^1}}{(\pi (2l-1)N)^p} 
        \end{split}
    \end{equation}
    Therefore, for any $p \geq d + \alpha + 2$, 
    \begin{equation}
        \begin{split}
            & \quad \left|\sum_{j\in \mathbb{Z}^d} 2^{\alpha}\pi^{\alpha}\|m+jN\|^{\alpha} \hat{u}_{m+jN}(t) -2^{\alpha}\pi^{\alpha}\|i(m)\|^{\alpha} \hat{u}_{i(m)}(t)\right| \\
            & \leq \left(\sum_{l=1}^{\infty} \frac{((l+1)^d-l^d)(2l+1)^{\alpha}}{(2l-1)^p} \right)   \frac{d^{\alpha/2} \max_{j}\|\partial_{x_j}^p u(t,x) \|_{L^1}}{\pi^{p-\alpha} N^{p-\alpha}} \\
            & \leq \left((2^d-1)3^\alpha + \sum_{l=2}^{\infty} \frac{((l+1)^d-l^d)(2l+1)^{\alpha}}{(2l-1)^p} \right)   \frac{d^{\alpha/2}\max_{j}\|\partial_{x_j}^p u(t,x) \|_{L^1}}{\pi^{p-\alpha} N^{p-\alpha}} \\
            & \leq \left((2^d-1)3^\alpha + 2^{\alpha}\sum_{l=2}^{\infty} \frac{1}{(2l-1)^{p-d-\alpha}} \right)   \frac{d^{\alpha/2}\max_{j}\|\partial_{x_j}^p u(t,x) \|_{L^1}}{\pi^{p-\alpha} N^{p-\alpha}} \\
            & \leq \left((2^d-1)3^\alpha + 2^{\alpha}\frac{1}{2(p-d-\alpha-1)} \right)   \frac{d^{\alpha/2}\max_{j}\|\partial_{x_j}^p u(t,x) \|_{L^1}}{\pi^{p-\alpha} N^{p-\alpha}} \\
            & \leq \frac{2^d 3^{\alpha} d^{\alpha/2} \max_{j}\|\partial_{x_j}^p u(t,x) \|_{L^1}}{\pi^{p-\alpha} N^{p-\alpha}}. 
        \end{split}
    \end{equation}
    As a result, 
    \begin{equation}\label{eqn:eq2}
    \begin{split}
        & \quad \left\|(\mathcal{F}^{-1})^{\otimes d} \left((-\Delta)^{\alpha/2}u(t,n/N)\right)_{n\in[N]^d} - N^{d/2} \sum_{m\in [N]^d} 2^{\alpha}\pi^{\alpha}\|i(m)\|^{\alpha} \hat{u}_{i(m)}(t)  \ket{m_0}\cdots\ket{m_{d-1}}\right\| \\
         &= N^{d/2} \left\|\sum_{m\in [N]^d} \left(\sum_{j\in \mathbb{Z}^d} 2^{\alpha}\pi^{\alpha}\|m+jN\|^{\alpha} \hat{u}_{m+jN}(t) -2^{\alpha}\pi^{\alpha}\|i(m)\|^{\alpha} \hat{u}_{i(m)}(t)\right) \ket{m_0}\cdots\ket{m_{d-1}}\right\| \\
         & \leq \frac{2^d 3^{\alpha} d^{\alpha/2} \max_{j}\|\partial_{x_j}^p u(t,x) \|_{L^1}}{\pi^{p-\alpha} N^{p-d-\alpha}}. 
    \end{split}
    \end{equation}
    Combining~\cref{eqn:eq1} and~\cref{eqn:eq2}, we have 
    \begin{equation}
        \left\|D(\mathcal{F}^{-1})^{\otimes d} \vec{u}_e(t) - (\mathcal{F}^{-1})^{\otimes d} \left((-\Delta)^{\alpha/2}u(t,n/N)\right)_{n\in[N]^d} \right\| \leq \frac{2^{p+1} d^{\alpha/2} \max_{j}\|\partial_{x_j}^p u(t,x) \|_{L^1}}{\pi^{p-\alpha} N^{p-d-\alpha}},  
    \end{equation}
    and 
    \begin{equation}
        \|r(t)\| \leq  \frac{2^{p+1} d^{\alpha/2} \max_{j}\|\partial_{x_j}^p u(t,x) \|_{L^1}}{\pi^{p-\alpha} N^{p-d-\alpha}}. 
    \end{equation}
    Plugging this back into~\cref{eqn:discretization_error_intermediate} and we obtain 
    \begin{equation}
        \|\vec{u}_e(T) - \vec{u}(T)\| \leq T \frac{2^{p+1} d^{\alpha/2} \max_{t,j}\|\partial_{x_j}^p u(t,x) \|_{L^1}}{\pi^{p-\alpha} N^{p-d-\alpha}}. 
    \end{equation}
\end{proof}

\begin{proof}[Proof of~\cref{cor:FRDE_potential_spatial_error_exp}]
    According to~\cref{lem:FRDE_potential_spatial_error}, the spatial discretization error can be bounded by 
    \begin{equation}
    \begin{split}
        \|(u(T,n/N))_{n\in[N]^d} - \vec{u}(T)\| & \leq 
        T \frac{2^{p+1} d^{\alpha/2} \max_{t,j}\|\partial_{x_j}^p u(t,x) \|_{L^1}}{\pi^{p-\alpha} N^{p-d-\alpha}}\\
        &\leq 2 \Lambda \pi^{\alpha} d^{\alpha/2} T \frac{ (2\Lambda/\pi)^{p} (p!)^{\sigma} }{ N^{p-d-\alpha}} \\
        & \leq 2^{1+\sigma} \pi^{\alpha+\sigma/2} \Lambda d^{\alpha/2} T \frac{p^{\sigma/2} (2\Lambda/\pi)^{p}  (p/e)^{p\sigma} }{ N^{p-d-\alpha}}
    \end{split}
    \end{equation}
    where in the second line we use $p! \leq \sqrt{2\pi p} (p/e)^p e^{1/(12p)} \leq 2\sqrt{\pi p} (p/e)^p $. 
    We choose $p = \lfloor p_* \rfloor$ where 
    \begin{equation}
        p_* =  \left(\frac{\pi}{2\Lambda}\right)^{1/\sigma} N^{1/\sigma}. 
    \end{equation}
    We remark that here we need $p_* \geq d + \alpha + 2$ and thus $N \geq (2\Lambda / \pi)(d + \alpha + 2)^{\sigma} $. 
    Then, 
    \begin{equation}
        \begin{split}
            \|(u(T,n/N))_{n\in[N]^d} - \vec{u}(T)\|  & \leq 2^{2+\sigma} \pi^{\alpha+\sigma/2} \Lambda d^{\alpha/2} T \frac{p_*^{\sigma/2} (2\Lambda/\pi)^{p_*}  (p_*/e)^{p_*\sigma} }{ N^{p_*-d-\alpha}} \\
            & = 2^{3/2+\sigma} \pi^{\alpha+\sigma/2+1/2} \Lambda^{1/2} T d^{\alpha/2} N^{d+\alpha+1/2} \exp\left(- \sigma \left(\frac{\pi}{2\Lambda}\right)^{1/\sigma} N^{1/\sigma} \right). 
        \end{split}
    \end{equation}
    By the Taylor expansion $e^x = \sum_{k} \frac{1}{k!} x^k$ and thus $e^x \geq  \frac{1}{k!} x^k$ for $x \geq 0$, we have 
    \begin{equation}
    \begin{split}
        \exp\left( \frac{\sigma}{2} \left(\frac{\pi}{2\Lambda}\right)^{1/\sigma} N^{1/\sigma} \right) &\geq \frac{1}{ \lceil \sigma(d+\alpha+1/2) \rceil ! } \left(\frac{\sigma}{2} \left(\frac{\pi}{2\Lambda}\right)^{1/\sigma} \right)^{\lceil \sigma(d+\alpha+1/2) \rceil } N^{d+\alpha+1/2}  \\
        & \geq  \left(\frac{\sigma}{2\lceil \sigma(d+\alpha+1/2) \rceil} \left(\frac{\pi}{2\Lambda}\right)^{1/\sigma} \right)^{\lceil \sigma(d+\alpha+1/2) \rceil } N^{d+\alpha+1/2}. 
    \end{split}
    \end{equation}
    Therefore 
    \begin{equation}
        \begin{split}
            & \quad \|(u(T,n/N))_{n\in[N]^d} - \vec{u}(T)\| \\
            & \leq 2^{3/2+\sigma} \pi^{\alpha+\sigma/2+1/2} \Lambda^{1/2} T d^{\alpha/2}  \left(\frac{2\lceil \sigma(d+\alpha+1/2) \rceil}{\sigma} \left(\frac{2\Lambda}{\pi}\right)^{1/\sigma} \right)^{\lceil \sigma(d+\alpha+1/2) \rceil } \exp\left(- \frac{\sigma}{2} \left(\frac{\pi}{2\Lambda}\right)^{1/\sigma} N^{1/\sigma} \right) \\
            & \leq c_1 T (c_2d)^{c_3 d} d^{\alpha/2} \exp\left( -c_4 N^{1/\sigma} \right), 
        \end{split}
    \end{equation}
    where we may choose 
    \begin{align}
        c_1 &= 2^{3/2+\sigma} \pi^{\alpha+\sigma/2+1/2} \Lambda^{1/2}, \\
        c_2 &= \max\left\{2(2+\alpha+1/\sigma) \left(\frac{2\Lambda}{\pi}\right)^{1/\sigma}, 1\right\},\\
        c_3 &= \sigma(3+\alpha),\\
        c_4 &= \frac{\sigma}{2} \left(\frac{\pi}{2\Lambda}\right)^{1/\sigma}.  
    \end{align}
\end{proof}

\begin{proof}[Proof of~\cref{cor:FRDE_potential_spatial_N}]
     By~\cref{cor:FRDE_potential_spatial_error_exp} and~\cref{lem:error_vector_quantum_state}, we have 
     \begin{equation}
         \|\ket{(u(T,n/N))_{n\in[N]^d}} - \ket{\vec{u}(T)}\| \leq \frac{2c_1 T}{\|(u(T,n/N))_{n\in[N]^d}\|} (c_2d)^{c_3 d} d^{\alpha/2} \exp\left( -c_4 N^{1/\sigma} \right). 
     \end{equation}
     The choice of $N$ can be solved by letting the error bound smaller than $\epsilon$. 
\end{proof}

\section{Bounding the Trotter error}\label{app:Trotter_error}

In this section we present the detailed proof of~\cref{lem:FRDE_linear_potential_Trotter_error}. 
For $s \in [t_0,t_0+h]$ and $t\in[0,h]$, we consider three operators 
    \begin{equation}
        S(s+t,s) = \mathcal{T} e^{\int_s^{s+t} (-B-{C}(\tau))d\tau}, 
    \end{equation}
    \begin{equation}
        \widetilde{S}(s+t,s) = e^{(-B-{C}(t_0+h/2)) t}, 
    \end{equation}
    and 
    \begin{equation}
        S_2(s+t,s) = e^{-B t/2} e^{-{C}(t_0+h/2) t}  e^{-B t/2}. 
    \end{equation}
    The proof follows the steps that we first bound $\|S-I\|$ and $\|\widetilde{S}-I\|$, then separately bound $\|S-\widetilde{S}\|$ and $\|\widetilde{S}-S_2\|$ using variation of parameters formula.  

    \subsection{Local error due to the time dependency}
    First, $S(s+t,s)$ satisfies the differential equation 
    \begin{equation}\label{eqn:proof_diff_eq_S}
        \frac{d}{dt} S(s+t,s) = (-B-{C}(s+t) ) S(s+t,s). 
    \end{equation}
    Integrate this differential equation and get 
    \begin{equation}
        S(s+t,s) - I = \int_s^{s+t} (-B-{C}(s+\tau) ) S(s+\tau,s) d\tau, 
    \end{equation}
    and thus we may write 
    \begin{equation}\label{eqn:proof_eq_S_I}
        S(s+t,s) = I + t R_1(s+t,s)
    \end{equation}
    where 
    \begin{equation}
        \| R_1(s+t,s) \| \leq \left(\|B\| + \max_{\tau}\|{C}(\tau)\|\right). 
    \end{equation}
    Similarly, 
    \begin{equation}\label{eqn:proof_eq_St_I}
        \widetilde{S}(s+t,s) = I + t R_2(s+t,s)
    \end{equation}
    where 
    \begin{equation}
        \| R_2(s+t,s) \| \leq \left(\|B\| + \max_{\tau}\|{C}(\tau)\|\right). 
    \end{equation}
    
    Now we bound the distance between $S$ and $\widetilde{S}$. 
    Notice that $\widetilde{S}$ satisfies the differential equation 
    \begin{equation}
    \begin{split}
        \frac{d}{dt} \widetilde{S}(t_0+t,t_0) &= (-B-{C}(t_0+h/2))  \widetilde{S}(t_0+t,t_0) \\
        & = (-B-{C}(t_0+t))  \widetilde{S}(t_0+t,t_0) + ({C}(t_0+t) - {C}(t_0+h/2)) \widetilde{S}(t_0+t,t_0). 
    \end{split}
    \end{equation}
    By Taylor expansion, we can write 
    \begin{equation}
        {C}(t_0+t) = {C}(t_0+h/2) + (t-h/2){C}'(t_0+h/2) + R_3(t), 
    \end{equation}
    where 
    \begin{equation}
        \|R_3(t)\| \leq \frac{(t-h/2)^2}{2} \max_{\tau}\|{C}''(\tau)\|. 
    \end{equation}
    Therefore the differential equation becomes 
    \begin{equation}
        \frac{d}{dt} \widetilde{S}(t_0+t,t_0) = (-B-{C}(t_0+t))  \widetilde{S}(t_0+t,t_0) + ((t-h/2){C}'(t_0+h/2) + R_3(t)) \widetilde{S}(t_0+t,t_0). 
    \end{equation}
    Comparing this with~\cref{eqn:proof_diff_eq_S} and using~\cref{lem:variation_of_parameters}, we obtain 
    \begin{equation}
        \widetilde{S}(t_0+h,t_0) = S(t_0+h,t_0) + \int_{0}^{h} S(t_0+h,t_0+t) ((t-h/2){C}'(t_0+h/2) + R_3(t)) \widetilde{S}(t_0+t,t_0) dt. 
    \end{equation}
    Plugging in~\cref{eqn:proof_eq_S_I} and~\cref{eqn:proof_eq_St_I} yields 
    \begin{equation}
    \begin{split}
        \widetilde{S}(t_0+h,t_0) &= S(t_0+h,t_0) + \int_{0}^{h} (I+(h-t)R_1(t_0+h,t_0+t))(t-h/2){C}'(t_0+h/2) (I+tR_2(t_0+t,t_0))  dt \\
        & \quad\quad\quad\quad\quad + \int_{0}^{h} S(t_0+h,t_0+t) R_3(t)\widetilde{S}(t_0+t,t_0) dt \\
        &= S(t_0+h,t_0) + \int_{0}^{h} (t-h/2){C}'(t_0+h/2)  dt \\
        & \quad\quad\quad\quad\quad + \int_{0}^{h} (h-t)(t-h/2)R_1(t_0+h,t_0+t){C}'(t_0+h/2) \widetilde{S}(t_0+t,t_0)  dt \\
        & \quad\quad\quad\quad\quad + \int_{0}^{h} t(t-h/2){C}'(t_0+h/2) R_2(t_0+t,t_0) dt \\
        & \quad\quad\quad\quad\quad + \int_{0}^{h} S(t_0+h,t_0+t) R_3(t)\widetilde{S}(t_0+t,t_0) dt \\
        & = S(t_0+h,t_0) + \int_{0}^{h} (h-t)(t-h/2)R_1(t_0+h,t_0+t){C}'(t_0+h/2) \widetilde{S}(t_0+t,t_0)  dt \\
        & \quad\quad\quad\quad\quad + \int_{0}^{h} t(t-h/2){C}'(t_0+h/2) R_2(t_0+t,t_0) dt \\
        & \quad\quad\quad\quad\quad + \int_{0}^{h} S(t_0+h,t_0+t) R_3(t)\widetilde{S}(t_0+t,t_0) dt. 
    \end{split}
    \end{equation}
    Therefore 
    \begin{equation}\label{eqn:proof_Magnus_error}
        \begin{split}
            & \quad \|\widetilde{S}(t_0+h,t_0) - S(t_0+h,t_0)\| \\
            & \leq \int_{0}^{h}t|t-h/2|dt \left(\max(\|R_1\|\|{C}'\|) + \max(\|{C}'\|\|R_2\|)\right) + \int_{0}^{h} \|R_3(t)\| dt \\
            & \leq \int_{0}^{h}t|t-h/2|dt \left(\max(\|R_1\|\|{C}'\|) + \max(\|{C}'\|\|R_2\|)\right) + \int_{0}^{h} \frac{(t-h/2)^2}{2} dt \max\|{C}''\| \\
            & = h^3 \left(  \frac{1}{4}(\|B\|+\max\|{C}\|)\max\|{C}'\|  + \frac{1}{24}\max\|{C}''\| \right). 
        \end{split}
    \end{equation}

    \subsection{Local time-independent Trotter error}
    
    Next, we bound the distance between $\widetilde{S}$ and $S_2$. 
    For notation simplicity, throughout this subsection we let $A_1 = -B$ and $A_2 = -{C}(t_0+h/2)$. 
    Differentiating $\widetilde{S}(t_0+t,t_0)$ and $e^{A_1 (t-s)}$ yields 
    \begin{equation}
        \frac{d}{dt} \widetilde{S}(t_0+t,t_0) = A_1 \widetilde{S}(t_0+t,t_0) + A_2 \widetilde{S}(t_0+t,t_0), 
    \end{equation}
    and 
    \begin{equation}
        \frac{d}{dt} (e^{A_1 (t-s)}) = A_1 e^{A_1 (t-s)}. 
    \end{equation}
    By~\cref{lem:variation_of_parameters}, for any $t\in [0,h]$, we have 
    \begin{equation}
        \widetilde{S}(t_0+t,t_0) = e^{A_1 t } + \int_0^t e^{A_1 (t-s)} A_2 \widetilde{S}(t_0+s,t_0) ds. 
    \end{equation}
    We use this equation iteratively for three times and get 
    \begin{equation}
        \begin{split}
             \widetilde{S}(t_0+h,t_0) &= e^{A_1 h } + \int_0^h e^{A_1 (h-t)} A_2 \widetilde{S}(t_0+t,t_0) dt \\
             & = e^{A_1 h } + \int_0^h e^{A_1 (h-t)} A_2 e^{A_1 t } dt + \int_0^h \int_0^t e^{A_1 (h-t)} A_2 e^{A_1 (t-s)} A_2 \widetilde{S}(t_0+s,t_0) ds dt \\
             & = e^{A_1 h } + \int_0^h e^{A_1 (h-t)} A_2 e^{A_1 t } dt + \int_0^h \int_0^t e^{A_1 (h-t)} A_2  e^{A_1 (t-s)} A_2 e^{A_1 s } ds dt + R_4(h), 
        \end{split}
    \end{equation}
    where 
    \begin{equation}\label{eqn:proof_R4_def}
        R_4(h) = \int_0^h \int_0^t \int_0^s e^{A_1 (h-t)} A_2 e^{A_1 (t-s)} A_2 e^{A_1 (s-\tau)} A_2 \widetilde{S}(t_0+\tau,t_0) d\tau ds dt. 
    \end{equation}
    Meanwhile, using the Taylor expansion of $e^{A_2 h}$, we may obtain 
    \begin{equation}
    \begin{split}
        S_2(t_0+h,t_0) &= e^{A_1 h/2} e^{A_2 h} e^{A_1 h/2} \\
        & = e^{A_1 h}  + h e^{A_1 h/2} A_2 e^{A_1 h/2} + \frac{h^2}{2} e^{A_1 h/2} A_2^2 e^{A_1 h/2} + R_5(h), 
    \end{split}
    \end{equation}
    where 
    \begin{equation}\label{eqn:proof_R5_def}
        R_5(h) = e^{A_1 h/2} \left(\int_0^h \frac{(h-t)^2}{2} A_2^3 e^{A_2 t} dt \right) e^{A_1 h/2}. 
    \end{equation}
    Therefore, 
    \begin{equation}\label{eqn:proof_Trotter_error_intermediate}
        \begin{split}
            & \quad \widetilde{S}(t_0+h,t_0) - S_2(t_0+h,t_0) \\
            & = \left(\int_0^h e^{A_1 (h-t)} A_2 e^{A_1 t } dt - h e^{A_1 h/2} A_2 e^{A_1 h/2}\right) \\
            & \quad + \left(\int_0^h \int_0^t e^{A_1 (h-t)} A_2  e^{A_1 (t-s)} A_2 e^{A_1 s } ds dt - \frac{h^2}{2} e^{A_1 h/2} A_2^2 e^{A_1 h/2}\right) \\
            & \quad + R_4(h) - R_5(h). 
        \end{split}
    \end{equation}
    It remains to bound each term in the right hand side of~\cref{eqn:proof_Trotter_error_intermediate}. 
    Let $f(t) = e^{A_1 (h-t)} A_2 e^{A_1 t}$, then 
    \begin{equation}
        \begin{split}
            & \quad \int_0^h e^{A_1 (h-t)} A_2 e^{A_1 t } dt - h e^{A_1 h/2} A_2 e^{A_1 h/2} \\
            & = \int_0^h f(t) dt - h f(h/2) \\
            & = \int_0^h \left(f(h/2) + (t-h/2) f'(h/2) + \int_0^t (t-s) f''(s) ds \right)dt - h f(h/2) \\
            & = \int_0^h \int_0^t (t-s) f''(s) ds dt. 
        \end{split}
    \end{equation}
    Notice that $f''(t) = e^{A_1 (h-t)} [[A_2,A_1],A_1] e^{A_1 t}$, so we bound 
    \begin{equation}\label{eqn:proof_Trotter_error_eq1}
        \left\|\widetilde{S}(t_0+h,t_0) - S_2(t_0+h,t_0)\right\| \leq \int_0^h \int_0^t |t-s| ds dt \max_{\tau\in[0,h]} \|f''(\tau)\| \leq \frac{1}{6} h^3 \|[[A_2,A_1],A_1] \|. 
    \end{equation}
    Similarly, let $g(t,s) = e^{A_1 (h-t)} A_2  e^{A_1 (t-s)} A_2 e^{A_1 s }$, then 
    \begin{equation}
        \begin{split}
            & \quad \int_0^h \int_0^t e^{A_1 (h-t)} A_2  e^{A_1 (t-s)} A_2 e^{A_1 s } ds dt - \frac{h^2}{2} e^{A_1 h/2} A_2^2 e^{A_1 h/2} \\
            & = \int_0^h \int_0^t g(t,s) ds dt - \frac{h^2}{2} g(h/2,h/2) \\
            & = \int_0^h \int_0^t \left(g(h/2,h/2) + \int_0^1 \nabla g(h/2+\alpha(t-h/2),h/2+\alpha(s-h/2)) \cdot \left(t-h/2,s-h/2 \right) d\alpha \right) ds dt \\
            & \quad\quad\quad\quad - \frac{h^2}{2} g(h/2,h/2) \\
            & =  \int_0^h \int_0^t \int_0^1 \nabla g(h/2+\alpha(t-h/2),h/2+\alpha(s-h/2)) \cdot \left(t-h/2,s-h/2 \right) d\alpha  ds dt. 
        \end{split}
    \end{equation}
    For $0\leq \alpha \leq 1$, $0 \leq s \leq t$ and $0 \leq t \leq h$, we always have $0 \leq h/2+\alpha(s-h/2) \leq h/2+\alpha(t-h/2)$ and $0 \leq h/2+\alpha(t-h/2) \leq h$, so 
    \begin{equation}
        \begin{split}
            & \quad \left\|\int_0^h \int_0^t e^{A_1 (h-t)} A_2  e^{A_1 (t-s)} A_2 e^{A_1 s } ds dt - \frac{h^2}{2} e^{A_1 h/2} A_2^2 e^{A_1 h/2}\right\| \\
            & \leq \int_0^h\int_0^t \int_0^1 \left(|t-h/2|+|s-h/2|\right) d\alpha ds dt \left( \max\left\{ \max_{0\leq s'\leq t'\leq h} \left\|\frac{\partial g(t',s')}{\partial t}\right\|, \max_{0\leq s'\leq t'\leq h} \left\|\frac{\partial g(t',s')}{\partial s}\right\|  \right\}  \right) \\
            & = \frac{1}{4} h^3 \left( \max\left\{ \max_{0\leq s'\leq t'\leq h} \left\|\frac{\partial g(t',s')}{\partial t}\right\|, \max_{0\leq s'\leq t'\leq h} \left\|\frac{\partial g(t',s')}{\partial s}\right\|  \right\}  \right). 
        \end{split}
    \end{equation}
    We may compute that 
    \begin{equation}
        \frac{\partial g}{\partial t} = e^{A_1 (h-t)} [A_2,A_1]  e^{A_1 (t-s)} A_2 e^{A_1 s }, 
    \end{equation}
    and 
    \begin{equation}
        \frac{\partial g}{\partial s} = e^{A_1 (h-t)} A_2 e^{A_1 (t-s)} [A_2,A_1] e^{A_1 s }, 
    \end{equation}
    so for $0 \leq s \leq t \leq h$, both $\|\partial g/\partial t\|$ and $\|\partial g/\partial s\|$ are bounded by $\|[A_2,A_1]\|\|A_2\|$. 
    Then 
    \begin{equation}\label{eqn:proof_Trotter_error_eq2}
        \left\|\int_0^h \int_0^t e^{A_1 (h-t)} A_2  e^{A_1 (t-s)} A_2 e^{A_1 s } ds dt - \frac{h^2}{2} e^{A_1 h/2} A_2^2 e^{A_1 h/2}\right\| \leq \frac{1}{4}h^3 \|[A_2,A_1]\| \|A_2\|. 
    \end{equation}
    The bounds for $R_4$ and $R_5$ are straightforward from~\cref{eqn:proof_R4_def} and~\cref{eqn:proof_R5_def} that 
    \begin{equation}\label{eqn:proof_Trotter_error_eq3}
        \|R_4(h)\| \leq \frac{1}{6} h^3 \|A_2\|^3 
    \end{equation}
    and 
    \begin{equation}\label{eqn:proof_Trotter_error_eq4}
        \|R_5(h)\| \leq \frac{1}{6} h^3 \|A_2\|^3. 
    \end{equation}
    Plug~\cref{eqn:proof_Trotter_error_eq1,eqn:proof_Trotter_error_eq2,eqn:proof_Trotter_error_eq3,eqn:proof_Trotter_error_eq4} back to~\cref{eqn:proof_Trotter_error_intermediate}, and we get 
    \begin{equation}\label{eqn:proof_Trotter_error}
    \begin{split}
        \left\|\widetilde{S}(t_0+h,t_0) - S_2(t_0+h,t_0)\right\| &\leq h^3 \left( \frac{1}{6}\|[A_1,[A_1,A_2]]\| + \frac{1}{4} \|[A_1,A_2]\|\|A_2\| + \frac{1}{3} \|A_2\|^3 \right) \\
        & \leq h^3 \left( \frac{1}{6} \max\|[B,[B,{C}]]\| + \frac{1}{4} \max\|[B,{C}]\|\max\|{C}\| + \frac{1}{3} \max\|{C}\|^3 \right). 
    \end{split}
    \end{equation}
    
    \subsection{Global error}
    
    Combining~\cref{eqn:proof_Magnus_error} and~\cref{eqn:proof_Trotter_error}, we obtain the local error bound as 
    \begin{equation}
    \begin{split}
        & \quad \|S(t_0+h,t_0) - S_2(t_0+h,t_0)\| \\
        & \leq h^3 \left(  \frac{1}{24}\max\|{C}''\| +  \frac{1}{4}(\|B\|+\max\|{C}\|)\max\|{C}'\|\right. \\
        & \quad\quad\quad\quad\quad \left. + \frac{1}{6}  \max\|[B,[B,{C}]]\| + \frac{1}{4} \max\|[B,{C}]\|\max\|{C}\| + \frac{1}{3} \max\|{C}\|^3 \right). 
    \end{split}
    \end{equation}
    
    Finally we bound the global error. 
    Notice that, for any $t\geq 0$, $\|S_2(t+s,t)\| \leq 1$ and $\|S(s+t,t)\| \leq 1$, then the global error accumulates linearly as 
    \begin{equation}
        \begin{split}
            & \quad \|S(T,0) - \prod_{j=0}^{r-1} S_2((j+1)h,jh)\| \\
            &= \left\| \prod_{j=0}^{r-1} S((j+1)h,jh) - \prod_{j=0}^{r-1} S_2((j+1)h,jh) \right\| \\
            & \leq \sum_{k=0}^{r-1} \left\|  \prod_{j=k+1}^{r-1}S_2((j+1)h,jh) \left( S((k+1)h,kh) - S_2((k+1)h,kh) \right) \prod_{j=0}^{k-1} S((j+1)h,jh)   \right\| \\
            & \leq \sum_{k=0}^{r-1} \left\|   S((k+1)h,kh) - S_2((k+1)h,kh)  \right\| \\
            & \leq T h^2 \left(  \frac{1}{24}\max\|{C}''\| +  \frac{1}{4}(\|B\|+\max\|{C}\|)\max\|{C}'\|   \right. \\
            & \quad\quad\quad\quad\quad \left. + \frac{1}{6}  \max\|[B,[B,{C}]]\| + \frac{1}{4}  \max\|[B,{C}]\|\max\|{C}\| + \frac{1}{3} \max\|{C}\|^3 \right). 
        \end{split}
    \end{equation}

\section{Bounding the commutators}\label{app:commutator}

\begin{proof}[Proof of~\cref{lem:commutator}]
    We shall consider a fixed $t$ and omit the explicit time dependence in our notation. 
    We first consider $[B, {C}]$ and write 
    \begin{equation}
    \begin{split}
        \|[B, {C}]\| &= \|[(\mathcal{F})^{\otimes d}  D (\mathcal{F}^{-1})^{\otimes d}, {C}]\| \\
            &= \| (\mathcal{F})^{\otimes d} [D,(\mathcal{F}^{-1})^{\otimes d} {C}(\mathcal{F})^{\otimes d}](\mathcal{F}^{-1})^{\otimes d}  \| \\
            & = \|  [D,(\mathcal{F}^{-1})^{\otimes d} {C}(\mathcal{F})^{\otimes d}] \|. 
    \end{split}
    \end{equation}
    By the definition of the matrices, we have 
    \begin{equation}
        \begin{split}
           {C}(\mathcal{F})^{\otimes d} 
            &= \frac{1}{N^{d/2}} \sum_{j,k\in[N]^d} {c}(j/N) \omega_N^{j\cdot k} \ket{j_0\cdots j_{d-1}}\bra{k_0\cdots k_{d-1}}
        \end{split}
    \end{equation}
    and 
    \begin{equation}
        \begin{split}
            (\mathcal{F}^{-1})^{\otimes d}{C}(\mathcal{F})^{\otimes d} 
            & = \frac{1}{N^d}  \sum_{j,k\in[N]^d} \left( \sum_{l\in [N]^d} \omega_N^{-j\cdot l} {c}(l/N) \omega_N^{l\cdot k} \right) \ket{j_0\cdots j_{d-1}}\bra{k_0\cdots k_{d-1}} \\
            & = \frac{1}{N^d}  \sum_{j,k\in[N]^d} \left( \sum_{l\in [N]^d}  \left( \sum_{m\in \mathbb{Z}^d  } \hat{c}_m \omega_N^{l\cdot m} \right) \omega_N^{l\cdot (k-j)} \right) \ket{j_0\cdots j_{d-1}}\bra{k_0\cdots k_{d-1}} \\
            & = \frac{1}{N^d}  \sum_{j,k\in[N]^d} \left( \sum_{m\in \mathbb{Z}^d} \hat{c}_m \left( \sum_{l\in [N]^d}  \omega_N^{l\cdot (m-(j-k))} \right) \right) \ket{j_0\cdots j_{d-1}}\bra{k_0\cdots k_{d-1}} \\
            & =   \sum_{j,k\in[N]^d} \left( \sum_{m\in \mathbb{Z}^d} \hat{c}_{j-k + Nm}   \right) \ket{j_0\cdots j_{d-1}}\bra{k_0\cdots k_{d-1}}. 
        \end{split}
    \end{equation}
    Let $\hat{b}_l = \sum_{m\in \mathbb{Z}^d} \hat{c}_{l + Nm}$, and we can compute that 
    \begin{equation}
         [D,(\mathcal{F}^{-1})^{\otimes d} {C}(\mathcal{F})^{\otimes d}] = \sum_{j,k\in[N]^d} \left( \|i(j)\|^{\alpha} - \|i(k)\|^{\alpha} \right) \hat{b}_{j-k}  \ket{j_0\cdots j_{d-1}}\bra{k_0\cdots k_{d-1}}. 
    \end{equation}

    Notice the decomposition
    \begin{equation}
        \|i(j)\|^{\alpha} - \|i(k)\|^{\alpha} = (\|i(j)\|^{\alpha/2} - \|i(k)\|^{\alpha/2})^2 + 2(\|i(j)\|^{\alpha/2} - \|i(k)\|^{\alpha/2}) \|i(k)\|^{\alpha/2}. 
    \end{equation}
    We may write 
    \begin{equation}\label{eqn:proof_commutator_representation}
         [D,(\mathcal{F}^{-1})^{\otimes d} {C}(\mathcal{F})^{\otimes d}] = A_1 + 2 A_2 A_3
    \end{equation}
    where 
    \begin{equation}
        A_1 = \sum_{j,k\in[N]^d} \left( \|i(j)\|^{\alpha/2} - \|i(k)\|^{\alpha/2} \right)^2 \hat{b}_{j-k}  \ket{j_0\cdots j_{d-1}}\bra{k_0\cdots k_{d-1}}, 
    \end{equation}
    \begin{equation}
        A_2 = \sum_{j,k\in[N]^d} \left( \|i(j)\|^{\alpha/2} - \|i(k)\|^{\alpha/2} \right) \hat{b}_{j-k}  \ket{j_0\cdots j_{d-1}}\bra{k_0\cdots k_{d-1}}
    \end{equation}
    and 
    \begin{equation}
        A_3 = \sum_{k\in[N]^d} \|i(k)\|^{\alpha/2} \ket{k_0\cdots k_{d-1}}\bra{k_0\cdots k_{d-1}}. 
    \end{equation}
    Now we bound the norms of $A_j$'s. 
    The norm of $A_3$ is clearly 
    \begin{equation}\label{eqn:proof_commutator_norm_A3}
        \|A_3\| = \mathcal{O}(d^{\alpha/4}N^{\alpha/2}). 
    \end{equation}
    For $A_2$, let $q = \text{argmax} ||i(j_{q'})| - |i(k_{q'})||$.
    We use the inequality
    \begin{equation}\label{eqn:proof_commutator_intermediate_1}
    \begin{split}
        \left|\|i(j)\|^{\alpha/2} - \|i(k)\|^{\alpha/2} \right| &\leq \left| \|i(j)\| - \|i(k)\| \right|^{\alpha/2} \\
        & \leq \| |i(j)| - |i(k)| \|^{\alpha/2} \\
        & \leq d^{\alpha/4} ||i(j_{q})|-|i(k_{q})||^{\alpha/2}. 
    \end{split}
    \end{equation}
    Here the second line is the triangle inequality (and $|v|$ for a vector $v$ denotes its entrywise absolute), the third line is due to the definition of $q$, and the first line can be proved by the following arguments: let $b \geq a \geq 0$ be two real number and $f(x) = x^{\alpha/2}$, then $b^{\alpha/2}-a^{\alpha/2} = \int_a^b f'(x) dx \leq \int_0^{b-a} f'(x) dx = (b-a)^{\alpha/2}$ because $f'(x) = \frac{\alpha}{2} x^{\alpha/2-1}$ is monotonically decreasing for $0<\alpha \leq 2$. 
    Notice that the Fourier coefficient $|\hat{c}_l| = \mathcal{O}(|l_{q'}|^{-(2+d)})$ for any index $q'$, and when $|i(j_{q})-i(k_{q})| \leq N/2$ (\emph{c.f.}, $i(j_{q})-i(k_{q})>N/2$ or $i(j_{q})-i(k_{q}) < -N/2$), $\hat{c}_{i(j)-i(k)}$ (\emph{c.f.}, $\hat{c}_{i(j)-i(k)-N}$ or $\hat{c}_{i(j)-i(k)+N}$) is the term with lowest frequency in $\hat{b}_{j-k}$, 
    We have 
    \begin{equation}
    \begin{split}
        \sum_{k\in[N]^d} \left| \|i(j)\|^{\alpha/2} - \|i(k)\|^{\alpha/2} \right| |\hat{b}_{j-k}| &\leq d^{\alpha/4} \sum_{k\in[N]^d} ||i(j_{q})|-|i(k_{q})||^{\alpha/2} |\hat{b}_{j-k}| \\
        & \leq d^{\alpha/4} \sum_{k\in[N]^d,m\in[N]^d} \|j-k+Nm\|_{\infty}^{\alpha/2} |\hat{c}_{j-k+Nm}| \\
        & = \mathcal{O}(d^{\alpha/4}). 
    \end{split}
    \end{equation}
    Similarly, the summation over $j$ is also bounded by $\mathcal{O}(d^{\alpha/4})$. 
    These imply $\|A_2\|_1 = \mathcal{O}(d^{\alpha/4})$ and $ \|A_2\|_{\infty} = \mathcal{O}(d^{\alpha/4})$, which further implies that 
    \begin{equation}\label{eqn:proof_commutator_norm_A2}
        \|A_2\| = \mathcal{O}(d^{\alpha/4})
    \end{equation}
    since $\|A\|^2 \leq \|A\|_1\|A\|_{\infty}$ for any matrix $A$. 
    For $A_1$, we use~\cref{eqn:proof_commutator_intermediate_1} again that 
    \begin{equation}
        \left(\|i(j)\|^{\alpha/2} - \|i(k)\|^{\alpha/2} \right)^2 \leq d^{\alpha/2} ||i(j_{q})|-|i(k_{q})||^{\alpha}, 
    \end{equation}
    and 
    \begin{equation}\label{eqn:proof_commutator_norm_A1}
        \|A_1\| = \mathcal{O}(d^{\alpha/2})
    \end{equation}
    by the same argument for $A_2$ (with the only difference that we need the Fourier coefficient to decay as $|\hat{c}_{l}| = \mathcal{O}(|l_{q'}|^{-(3+d)})$). 
    Plugging~\cref{eqn:proof_commutator_norm_A1,eqn:proof_commutator_norm_A2,eqn:proof_commutator_norm_A3} back to~\cref{eqn:proof_commutator_representation} yields 
    \begin{equation}
        \|[B, {C}]\| = \mathcal{O}(d^{\alpha/2}N^{\alpha/2}). 
    \end{equation}
    
    The second estimate $\|[B,[B,{C}]]\| = \mathcal{O}(d^{\alpha} N^{\alpha})$ can be proved in the same way by using $|\hat{c}_{l}| = \mathcal{O}(|l_{q'}|^{-(5+d)})$ and noticing that 
    \begin{equation}
        \|[B,[B,{C}]]\| = \|[D,[D,(\mathcal{F}^{-1})^{\otimes d} {C}(\mathcal{F})^{\otimes d}]]\|, 
    \end{equation}
    \begin{equation}
        [D,[D,(\mathcal{F}^{-1})^{\otimes d} {C}(\mathcal{F})^{\otimes d}]] = \sum_{j,k\in[N]^d} \left( \|i(j)\|^{\alpha} - \|i(k)\|^{\alpha} \right)^2 \hat{b}_{j-k}  \ket{j_0\cdots j_{d-1}}\bra{k_0\cdots k_{d-1}}, 
    \end{equation}
    and 
    \begin{equation}
    \begin{split}
        & \quad (\|i(j)\|^{\alpha} - \|i(k)\|^{\alpha})^2 \\
        &= (\|i(j)\|^{\alpha/2} - \|i(k)\|^{\alpha/2})^4 + 4(\|i(j)\|^{\alpha/2} - \|i(k)\|^{\alpha/2})^3 \|i(k)\|^{\alpha/2} + 4(\|i(j)\|^{\alpha/2} - \|i(k)\|^{\alpha/2})^2 \|i(k)\|^{\alpha}. 
    \end{split}
    \end{equation}
\end{proof}

\section{Discretization error in the LCHS-IP method}\label{app:LCHS_discretization_error}

\begin{proof}[Proof of~\cref{lem:LCHS_discretization_error}]
    Let $U(t;\xi) = \mathcal{T} e^{-i \int_0^t  \xi(B+{C}(s)) ds } = e^{-i\xi B t} \mathcal{T} e^{-i \int_0^t H_I(s;\xi) ds}$ and $V(t;\xi) = \frac{1}{\pi(1+\xi^2)} U(t;\xi)$. 
    The truncation error can be bounded by 
    \begin{equation}\label{eqn:LCHS_med_truncation_error}
        \left\| \mathcal{T} e^{ - \int_0^T (B+{C}(s)) ds } - \int_{-\Xi}^{\Xi} \frac{1}{\pi(1+\xi^2)}U(T;\xi) d\xi \right\| \leq 2 \int_{\Xi}^{\infty} \frac{d\xi}{\pi(1+\xi^2)} = \frac{2}{\pi}(\frac{\pi}{2} - \arctan(\Xi) ) \leq \frac{2}{\pi \Xi}. 
    \end{equation}
    According to the standard quadrature error bound, we have 
    \begin{equation}\label{eqn:LCHS_med_quadrature_error}
        \left\| \mathcal{T} e^{ - \int_0^T (B+{C}(s)) ds } - \sum_{j=0}^{M-1} w_j U(T;\xi_j) \right\| \leq \frac{2\Xi^2}{M} \sup_{\xi\in[-\Xi,\Xi]} \left\| \frac{\partial V}{\partial \xi}\right\|. 
    \end{equation}
    To estimate $\partial V/\partial \xi$, we first compute $\partial U/\partial \xi$. 
    Differentiating~\cref{eqn:LCHS_def_U_ODE} yields 
    \begin{equation}
        \frac{\partial}{\partial t} \frac{\partial U}{\partial \xi} = -i(B+{C}(t)) U - i (\xi B+ \xi{C}(t)) \frac{\partial U}{\partial \xi}, \quad \frac{\partial U}{\partial \xi}(0;\xi) = 0. 
    \end{equation}
    By the variation of constants formula, we have 
    \begin{equation}
        \frac{\partial U}{\partial \xi}(t;\xi) = \int_0^t \left(\mathcal{T} e^{-i\int_s^t (\xi B+\xi{C}(\tau)) d\tau }   \right) (-i)(B+{C}(s)) U(s;\xi) ds, 
    \end{equation}
    and thus 
    \begin{equation}
        \left\| \frac{\partial U}{\partial \xi} \right\| \leq T (\|B\|+\max\|{C}\|). 
    \end{equation}
    Therefore, by the product rule, we have 
    \begin{equation}
    \begin{split}
        \left\| \frac{\partial V}{\partial \xi}\right\| &\leq \frac{2\pi|\xi|}{\pi^2(1+\xi^2)} + \frac{1}{\pi(1+\xi^2)} \left\|\frac{\partial U}{\partial \xi}\right\| \\
        & \leq \frac{1}{\pi} \left( 1 + T (\|B\|+\max\|{C}\|) \right). 
    \end{split}
    \end{equation}
    Plugging this back into~\cref{eqn:LCHS_med_quadrature_error} and~\cref{eqn:LCHS_med_truncation_error} gives the desired error bound, and the choices of $\Xi$ and $M$ directly follow from the error bound. 
\end{proof}

\end{document}